\pdfoutput=1

\documentclass[12pt]{article}
\usepackage{amsmath,epsf,amssymb,latexsym,cite,shadow,eucal,theorem,enumerate}
\usepackage[matrix,arrow,frame,import,curve,color]{xy}

\usepackage{graphicx}

\newtheorem{theorem}{Theorem}

{\theorembodyfont{\rmfamily}}

\newif\iffigs\figstrue

\DeclareFontFamily{U}{rsf}{}
\DeclareFontShape{U}{rsf}{m}{n}{
  <5> <6> rsfs5 <7> <8> <9> rsfs7 <10-> rsfs10}{}
\DeclareMathAlphabet\Scr{U}{rsf}{m}{n}

\def\O{\Scr{O}}
\def\C{{\mathbb C}}
\def\P{{\mathbb P}}

\def\Q{{\mathbb Q}}
\def\R{{\mathbb R}}
\def\Z{{\mathbb Z}}

\def\Im{\operatorname{Im}}
\def\Image{\operatorname{Image}}
\def\Hom{\operatorname{Hom}}

\def\Ext{\operatorname{Ext}}

\def\coker{\operatorname{coker}}

\def\Vol{\operatorname{Vol}}

\def\SU{\operatorname{SU}}
\def\GU{\operatorname{U{}}}

\def\rank{\operatorname{rank}}

\def\Cone{\operatorname{Cone}}
\def\Obj{\operatorname{Obj}}

\def\ch{\operatorname{\mathit{ch}}}
\def\td{\operatorname{\mathit{td}}}

\def\id{{\mathbf{1}}}

\def\p{\partial}

\def\CY{Calabi--Yau}
\def\LG{Landau--Ginzburg}

\def\GLSM{gauged linear $\sigma$-model}

\def\cP{{\Scr P}}

\def\cL{{\Scr L}}
\def\cE{{\Scr E}}
\def\cF{{\Scr F}}
\def\cX{{\Scr X}}

\def\DC{\mathbf{D}}

\def\ff#1#2{{\textstyle\frac{#1}{#2}}}

\def\mf#1{\mathfrak{#1}}

\def\pz{\phantom{-}}

\def\eqn#1#2{\begin{equation}#2
  \ifx{#1}{}\else\label{#1}\fi\end{equation}}

\def\sOX{\mathsf{O}_X}

\begin{document}

\begin{titlepage}
\begin{flushright}
September 2009
\end{flushright}
\vspace{.5cm}
\begin{center}
\baselineskip=16pt
{\fontfamily{ptm}\selectfont\bfseries\huge
Decompactifications and Massless D-Branes \\in Hybrid Models\\[20mm]}
{\bf\large  Paul S.~Aspinwall and M.~Ronen Plesser
 } \\[7mm]

{\small

Center for Geometry and Theoretical Physics, 
  Box 90318 \\ Duke University, 
 Durham, NC 27708-0318 \\ \vspace{6pt}

 }

\end{center}

\begin{center}
{\bf Abstract}
\end{center}
A method of determining the mass spectrum of BPS D-branes in any phase
limit of a gauged linear sigma model is introduced. A ring associated
to monodromy is defined and one considers K-theory to be a module over
this ring. A simple but interesting class of hybrid models with
Landau--Ginzburg fibres over $\P^n$ are analyzed using special
K\"ahler geometry and D-brane probes. In some cases the
hybrid limit is an infinite distance in moduli space and corresponds
to a decompactification. In
other cases the hybrid limit is at a finite distance and
acquires massless D-branes. An example studied
appears to correspond to a novel theory of supergravity with an
$\SU(2)$ gauge symmetry where the gauge and gravitational couplings
are necessarily tied to each other.


\end{titlepage}

\vfil\break


\section{Introduction}    \label{s:intro}

A huge class of $N=(2,2)$ superconformal field theories are associated
with the gauged linear $\sigma$-model of Witten \cite{W:phase}.  This
is an asymptotically free gauge theory which flows at low energies to
a nontrivial superconformal field theory.  The parameters of the
\GLSM\ include the coefficients $r$ of Fayet-Iliopoulos $D$-terms, and
for large $|r|$ the gauge symmetry is broken at high energies, where a
perturbative treatment of the gauge interaction is valid.  It is in
these asymptotic regions that the \GLSM\ is a reliable guide to the
low-energy physics.  As is well-known, this model exhibits a ``phase''
structure, dividing $r$-space into cones.  In each of these the space
of classical vacua is naturally modeled by a specific toric geometry;
this description is expected to be reliable deep in the interior of
each cone.  The significance of the limit points is especially clear
in the topological twisted version of the theory.  Here each limit
point is the center of an instanton expansion.  The most familiar
examples are \CY\ phases and \LG\ limit points.
 
If one considers an example where the moduli space has more than a few
dimensions, there will be hundreds of phases and most of these phases
will not have a simple interpretation as a \CY\ space (with partially
resolved orbifolds) or a \LG\ theory.  In a typical phase the space of
classical vacua will be reducible, including components of different
dimensions.  In some of these vacua there will be massless fields in
addition to the coordinates on the space of vacua.  These interact via
an effective superpotential.  Each irreducible component can na\"\i
vely be thought of as a fibration over some geometric space where the
fibre is a \LG\ theory and these phases are known as ``hybrid'' phases
\cite{AGM:I,W:phase}.  Despite their ubiquity, hybrid phases have not
been studied as thoroughly as the types of phases mentioned above.

The purpose of this paper is to explore the geometry of a few simple
hybrid phases by using D-brane probes to see how accurate the above
fibration picture is. In particular we want to study the geometry in
the ``limit'' of such hybrid phases, deep in the interior of the
associated cone.  In this limit the gauge symmetry is broken at high
energy and a classical calculation shows that the base of the
fibration (the space of classical vacua) grows large, meaning the
massless fields corresponding to motion along the base appear weakly
coupled.  

We find two types of behavior in the hybrid limit.  In hybrid phases
of the first type, which we will call {\em true hybrid\/}  phases, the
base of the fibration indeed grows large, and the description of the
theory as a \LG\ model fibered over the base is valid.   The hybrid
limit point, corresponding to a decompactification of the target
space, is at an infinite distance (in the Zamolodchikov metric) in
moduli space.  The gauge instanton expansion about this limit is an
expansion in worldsheet instantons wrapping holomorphic curves in the
base space.

In hybrid phases of the second type, which we call {\em
  pseudo-hybrid\/} phases, the description of the low-energy limit as
a fibration is not valid.  The base space does not grow large, but
remains at a finite size in the limit.  These limit points are at a
finite distance in the moduli space, and instead of a weakly-coupled
decompactification limit we find that they describe singular conformal
field theories.  The singularity is associated in string theory to the
appearance of massless D-branes at the hybrid limit. The spacetime
physics near such a singularity is an interesting strongly-interacting
theory with some unfamiliar properties.  
 
A central mathematical problem in this paper is the determination of
the asymptotic form of D-brane masses as we go into the hybrid
limit. This can be solved by directly solving the associated
Picard--Fuch's differential equations from mirror symmetry. However,
rather than doing this we find it much more enlightening to introduce
a construction we call the ``monodromy ring''. The ring arises from an
observation in \cite{HHP:linphase} that the monodromy on D-branes
around phase limits has a particularly simple form when the category
of D-branes is written in terms of specific tilting collections. By
using the monodromy ring we can, with very little effort, determine
the asymptotic form of the relevant periods in phase limits.

The monodromy ring results also allow us to easily construct
candidates for particular D-branes with a corresponding class
associated to a given period. By doing this we can find candidate
``0-branes''. That is, D-branes naturally associated with points in
the hybrid geometry. Such 0-branes are not at all the same D-branes
that would be called 0-branes in the large radius limit. These
0-branes allow us to probe the geometry of the hybrid phases. A
defining property of these 0-branes is that they are {\em light},
i.e., lighter than 2-branes, 4-branes, etc., as one approaches the phase
limit.

The geometry of hybrid models has been studied elsewhere, in
\cite{AG:gmi,Caldararu:2007tc}, for example. The work of 
\cite{Caldararu:2007tc} using twisted derived categories seems
important for the limits associated to a decompactification but uses
a somewhat different perspective than we adopt in this paper.

In section \ref{s:mon} we describe the mathematical machinery of the
monodromy ring.  Section \ref{s:decomp} discusses the two simplest
cases of the hybrid model and we see how they correspond to
decompactifications in, respectively, one and three complex
dimensions. In section \ref{s:compact} we discuss another very simple
hybrid model but discover that this does not correspond to a
decompactification but, rather, a curious model of $N=2$ supergravity
with an $\SU(2)$ gauge symmetry that is somehow unavoidably
intertwined with gravity.


\section{Monodromy} \label{s:mon}

We will need to compute the asymptotic form of periods for D-branes
near limit points in the moduli space of gauged linear
$\sigma$-models. To do this we use the monodromy ring defined
below. The mathematics of computing periods is well-established
beginning with the work of \cite{CDGP:} and consists of solving
Picard--Fuchs equations, say in the large radius phase, and
analytically continuing into the other phases. Here we develop a
somewhat easier method which should become particularly useful when
applied to cases of more than one modulus. In this paper we consider
only moduli spaces of dimension one and the periods can be obtained
fairly painlessly without resorting to the monodromy ring. Having said
that, the monodromy ring analysis lends itself well to finding
specific D-branes for our candidate 0-branes rather than just D-brane
charges.

\subsection{Noncompact \CY s} \label{ss:nonc}

While the main object of interest in this paper is a compact \CY\
threefold we will first discuss the simpler case where the \CY\ is
toric and noncompact.  We refer to \cite{me:toricD} for more details of
some of the constructions below.

Assume we have a noncompact toric \CY\ threefold $X$. This has a
homogeneous coordinate ring \cite{Cox:} $S=\C[x_1,x_2,x_3,\ldots,
x_n]$. There is a fan $\Sigma$ and each homogeneous coordinate $x_i$
is associated with one of the one-dimensional rays of $\Sigma$. The
Cox ideal in $S$ is defined as
\begin{equation}
  B_\Sigma =
  \left(\prod_{i\not\in\sigma_1}x_i,\prod_{i\not\in\sigma_2}x_i,\ldots
  \right),
\end{equation}
where $i\not\in\sigma$ means that the ray $i$ is not in a cone
$\sigma$ of the fan $\Sigma$. We also have an $r\times n$ charge
matrix $Q$ which defines a $(\C^*)^r$ torus action:
\begin{equation}
  x_i \mapsto \lambda_1^{Q_{1i}}\lambda_2^{Q_{2i}}\ldots
              \lambda_r^{Q_{ri}} x_i, \label{eq:Qs}
\end{equation}
where $\lambda_j\in\C^*$. 
This gives $S$ the structure of an $r$-graded ring.
The toric variety $X$ is then defined as the
quotient $(\C^n-V(B_\Sigma))/(\C^*)^r$, where $V(B_\Sigma)$ is the
collection of points where all elements of the ideal $B_\Sigma$
vanish.

The category of B-type D-branes on $X$ is then given
\cite{Cox:,HHP:linphase,me:toricD} by the quotient:
\begin{equation}
  \DC(X) = \frac{\DC(\textrm{gr-}S)}{T_\Sigma},  \label{eq:DCXq}
\end{equation}
where $\DC(\textrm{gr-}S)$ is the bounded derived category of finitely
generated graded $S$-modules and $T_\Sigma$ is the full subcategory of
modules killed by some power of the Cox ideal $B_\Sigma$. This
quotient of triangulated categories is as in \cite{BO:flop}.

The associated gauged linear $\sigma$-model has chiral multiplets
$\Phi_1,\Phi_2,\Phi_3,\ldots,\Phi_n$ interacting with $r$ gauge
multiplets. The gauge group acts as in (\ref{eq:Qs}) with
$|\lambda|=1$.  $X$ is \CY\ if the charges satisfy 
\begin{equation}
\sum_i Q_i=0 \label{eq:cycond}
\end{equation}
and this ensures the existence of a non-anomalous $R$-symmetry in the
theory which descends to the $\GU(1)_R$ of the low-energy superconformal
field theory.\footnote{The noncompact field space means this theory
  requires boundary conditions to be specified for the fields.  In
  practice, we consider this theory either as a local description near
  a singularity in a compact \CY\ or as a step in constructing a
  compact model as in section \ref{s:ccy}.}  There are $r$
Fayet-Iliopoulos couplings and a cone in $\R^r$ such that for values
of the couplings in this cone, the $D$-term
conditions constrain the expectation values of the scalars $x_i$ in
the chiral multiplets to lie away from $V(B_\Sigma)$.  The space of
gauge-inequivalent vacua is then precisely $X$.

In all of our examples here $r=1$ and there will be one coupling
conventionally denoted $r$.  We choose the charges so that the \CY\ 
cone is given by $r>0$.  The fields $x_i$ with negative
charge are relabeled $p_a$.  For these values of $r$ the space of gauge
inequivalent vacua is the total space $X$ of a smooth bundle over a
compact toric variety for which $x_i$ generate the homogeneous
coordinate ring.  For $r<0$ the space of vacua will
typically be a $V$-bundle over a toric variety for which $p_a$
generate the homogeneous coordinate ring.

\subsubsection{The monodromy ring} \label{sss:monring}

An interesting observation in \cite{HHP:linphase} is that the
monodromy of D-branes around the linear $\sigma$-model limit
associated to $\Sigma$ has a very simple description in terms of 
(\ref{eq:DCXq}). It amounts to a simple shift in the grading.

Let us introduce a construction for the associated K-theory of the
D-branes.  Consider first the K-theory of the category
$\DC(\textrm{gr-}S)$. This is the free abelian group generated all
objects in $\DC(\textrm{gr-}S)$ divided out by the relation
$[\mathsf{a}] - [\mathsf{b}] + [\mathsf{c}]=0$ for any distinguished
triangle
\begin{equation}
\xymatrix{& \mathsf{c}\ar[dl]|{[1]}&\\
\mathsf{a}\ar[rr]&&\mathsf{b}\ar[ul]
}
\end{equation}
Any object in $\DC(\textrm{gr-}S)$ has a finite resolution in terms of
the shifted modules $S(\mathbf{v})$ where $\mathbf{v}$ is an
$r$-component vector of integers. There are no nontrivial triangles
with entries purely of the form $S(\mathbf{v})$ and so the K-theory
group, $K(\textrm{gr-}S)$, of $\DC(\textrm{gr-}S)$ is isomorphic to
the abelian group of bounded Laurent polynomials in $r$
variables. That is, we have an isomorphism
\begin{equation}
\xymatrix@1{
k:K(\textrm{gr-}S)\ar[r]^-{\simeq}&
\Z[s_1,s_1^{-1},s_2,s_2^{-1},\ldots,s_r,s_r^{-1}],
}
\end{equation}
where $k(S(\mathbf{v}))=s_1^{v_1}s_2^{v_2}\cdots s_r^{v_r}$.

The K-theory of the derived category $\DC(X)$ is the usual topological
K-theory $K(X)$ if $X$ is a manifold (and we may use this as the
definition of $K(X)$ if $X$ is not a manifold) \cite{me:TASI-D}.

In the quotient (\ref{eq:DCXq}), the K-theory of the subcategory
$T_\Sigma$ will correspond to a subgroup of
$\Z[s_1,s_1^{-1},s_2,s_2^{-1},\ldots,s_r,s_r^{-1}]$. Since
$T_\Sigma$ is closed under shifting the grading, this subgroup will
actually be an ideal. We therefore have
\begin{equation}
  K(X) = \frac{\Z[s_1,s_1^{-1},s_2,s_2^{-1},\ldots,s_r,s_r^{-1}]}
              {k(T_\Sigma)}. \label{eq:KX}
\end{equation}

We may locally view a punctured neighborhood of any limit point in
the moduli space as
$(\C^*)^r$. According to \cite{HHP:linphase}, the monodromy around this
limit point within the $i$th copy of $\C^*$ may be obtained by
shifting the $i$th grading by one.  That is, in terms of
(\ref{eq:KX}), such monodromy amounts to multiplication by
$s_i$. This puts a ring structure on $K(X)$. We will refer to this
as the {\em monodromy ring}. 

In the large radius limit monodromy amounts to tensoring by a line
bundle $\cL$. In K-theory this corresponds to a wedge product by the
Chern character $\ch(\cL)$. This gives a direct link in the large
radius phase between the monodromy ring and the {\em cohomology ring}. Such
a link is less clear in the other phases.

\subsubsection{An example --- $\C^3/\Z_3$}

All of this is most easily understood in the context of an example so let us
consider the case of the blow-up of $\C^3/\Z_3$. Here we use the notation 
\begin{equation}
  S = \C[p,x_1,x_2,x_3],
\end{equation}
and the charges are $(-3,1,1,1)$ respectively.

In the orbifold phase $\Sigma$ consists of a single cone and
$B_\Sigma=(p)$. Thus, in the D-brane category $\DC(X)$, the module
$S/(p)$ is trivial. This module is quasi-isomorphic to the complex
\begin{equation}
\xymatrix@1{
  S(3) \ar[r]^-p& S}.
\end{equation}
This means that the map $p$ above is an {\em isomorphism\/} in the
quotient category $\DC(X)$ given by (\ref{eq:DCXq}). Similarly, since
$S(m)/(p)$ is trivial for any $m$ we have isomorphisms of D-branes
\begin{equation}
  S(m) \cong S(m+3),\label{eq:m3}
\end{equation}
in the orbifold phase. 

Monodromy around the limit point, i.e., the orbifold point, is simply
given by shifting the grading by one. Thus the equivalence
(\ref{eq:m3}) shows immediately that there is a $\Z_3$ monodromy
around the orbifold point in moduli space.

The other phase is the large radius limit phase where
$B_\Sigma=(x_1,x_2,x_3)$. Here the monodromy is not finite. To
understand the monodromy we can look at D-brane charges or, equivalently,
K-theory. That is, we look at the monodromy ring. We have a resolution
\begin{equation}
\xymatrix@C=12mm{
  0\ar[r]&S(-3)\ar[r]^-{\left(
         \begin{smallmatrix}x_1\\x_2\\x_3\end{smallmatrix}\right)}&
  S(-2)^{\oplus3}\ar[rr]^-{\left(\begin{smallmatrix}
    0&x_3&-x_2\\-x_3&0&x_1\\x_2&-x_1&0\end{smallmatrix}\right)}&&
  S(-1)^{\oplus3}\ar[r]^-{\left(\begin{smallmatrix}
         x_1&x_2&x_3\end{smallmatrix}\right)}&
  S\ar[r]&\frac{S}{(x_1,x_2,x_3)}\ar[r]&0
}
\end{equation}
The ideal $k(T_\Sigma)$ must therefore contain
\begin{equation}
  1 - 3s^{-1} + 3s^{-2} - s^{-3} = (1-s^{-1})^3.
\end{equation}
In fact the ideal $k(T_\Sigma)$ is generated by this. 
We denote this principal ideal as $\left((1-s^{-1})^3\right)$, which
is equivalent to $\left((s-1)^3 \right)$, since $s$ is a unit.
Thus the
monodromy ring is
\begin{equation}
  K(X) = \frac{\Z[s,s^{-1}]}{\left( (s-1)^3\right)}.
\end{equation}
This makes it clear that the monodromy action on D-brane charges is
unipotent of order 3, i.e., it obeys $(s-1)^3=0$.

Clearly in the orbifold phase we have a monodromy ring
\begin{equation}
  K(X) = \frac{\Z[s,s^{-1}]}{\left( s^3-1\right)}.
\end{equation}

It is worth noting that we are discussing the K-theory finitely
generated by vector bundles on $X$. This means that we actually miss
the D-brane charge of a 0-brane. Writing a free resolution of a
skyscraper sheaf, $\O_x$, associated to a point in $X$ one can show
that $k(\O_x)=0$. This problem is caused by the fact that $X$ is
noncompact. Fortunately the K-theory captures all of the
D-branes charges once we consider compact examples.


\subsection{Compact \CY s}   \label{s:ccy}

\subsubsection{The category of D-branes}

Our main interest is in compact \CY\ spaces.  To construct a gauged
linear $\sigma$-model whose low-energy physics is associated to a
compact \CY\ spaces \cite{W:phase} we introduce into the model
described above a superpotential interaction determined by a
gauge-invariant polynomial in the fields which we assume is of the
form
\begin{equation}
 \mathcal
  {W} = \sum_a p_a f_a(x) \label{eq:lrw}
\end{equation} 
where $f_a$ are homogeneous polynomials in $x_i$ such that $\mathcal
{W}$ is gauge invariant.  The low-energy dynamics in the large-$r$
limit is that of a nonlinear $\sigma$-model on the critical point set
of $\mathcal{W}$. For sufficiently generic $f_a$ this is the
intersection of all the hypersurfaces $f_a=0$ and $p_a=0$ in $V$. The
condition (\ref{eq:cycond}) together with the fact that all the
$x_i$'s have positive charge ensure that $X$ is a compact \CY\ variety.
The $\GU(1)_R$ symmetry is broken by the superpotential, but combining
this with the global (non-$R$) symmetry assigning charge 2 to $p_a$
and charge 0 to $x_i$, under which $\mathcal W$ has charge 2, produces
an unbroken symmetry which descends again to the $\GU(1)_R$ symmetry
of the superconformal field theory.

Following our discussion of the noncompact case, let
$S$ denote the homogeneous coordinate ring of $V$. $S$
has an $r$-fold grading from the toric data as at the start of this
section. In addition we add one further grading associated to the
$R$-symmetry. This grading lives in $2\Z$, i.e., it is
always an even number. 

Define the category $\textrm{DGr}S(\mathcal{W})$ of matrix
factorizations of $\mathcal{W}$ as follows. An object is a pair
\begin{equation}
  \bar P = \Bigl(
\xymatrix@1{
  P_1 \ar@<0.6mm>[r]^{u_1}&P_0\ar@<0.6mm>[l]^{u_0}
}\Bigr), \label{eq:Pdef}
\end{equation}
where $P_0$ and $P_1$ are two finite rank graded free $S$-modules.
The two maps satisfy the matrix factorization condition
\begin{equation}
  u_0u_1 = u_1u_0 = \mathcal{W}.\id, \label{eq:mfact}
\end{equation}
and $u_0$ is a map of degree 2 with respect to the $R$-symmetry while
$u_1$ has degree 0. Both $u_0$ and $u_1$ have degree 0 with respect to
the toric gradings. Morphisms are defined in the obvious way up to
homotopy. We refer to \cite{Orlov:mfc,me:csalg} for more details.
The category $\textrm{DGr}S(\mathcal{W})$ is a triangulated category
with a shift functor
\begin{equation}
  \bar P[1] = \Bigl(
\xymatrix@1{
  P_0 \ar@<0.6mm>[r]^{u_0}&P_1\{2\}\ar@<0.6mm>[l]^{u_1}
}\Bigr), \label{eq:shift1}
\end{equation}
where $\{\}$ denotes a shift in the $R$-grading. Thus
\begin{equation}
  \bar P[2] = \bar P\{2\},
\end{equation}
and the $R$-symmetry grading is identified with the homological
grading (and extended from $2\Z$-valued to $\Z$-valued). That is, there
is no difference between $[m]$ and $\{m\}$ in this category although
we will sometimes use both notations for clarity.

The category $\textrm{DGr}S(\mathcal{W})$ should be considered to be
the category of D-branes before the conditions imposed by the $D$-terms
of the gauged linear $\sigma$-model have been imposed. It is the
analogue of $\DC(\textrm{gr-}S)$ for the noncompact \CY\ case
considered above. 

There is an equivalence of categories \cite{Orlov:mfc}
\begin{equation}
  \mathrm{DGr}S(\mathcal{W})\cong 
      \frac{\DC(\textrm{gr-}S')}{\mf{Perf}(S')},  \label{eq:q8}
\end{equation}
where $S'=S/(\mathcal{W})$ and $\mf{Perf}(S')$ denotes the subcategory
of ``perfect complexes'', i.e., finite length complexes of
finitely-generated modules. This equivalence is made manifest by an
observation by Eisenbud \cite{Eis:mf}. Any free $S$-resolution of an
$S$-module annihilated by $\mathcal{W}$ will ultimately becomes
periodic of length 2 as one moves sufficiently far left in the
resolution. This asymptotic form of the resolution can be written as a
matrix factorization (\ref{eq:mfact}).

The actual category of D-branes is again a quotient
\begin{equation}
  \frac{\textrm{DGr}S(\mathcal{W})}{T_\Sigma},  \label{eq:comD}
\end{equation}
where $T_\Sigma$ is a subcategory of $\textrm{DGr}S(\mathcal{W})$
associated,
via (\ref{eq:q8}),
to modules annihilated by a power of the Cox ideal $B_\Sigma$.

\subsubsection{An example --- the quintic threefold}  \label{sss:quin1}

The previous discussion can be understood more easily in the context
of an example. Let $V$ be the total space of the canonical line bundle
over $\P^4$. We have a homogeneous coordinate ring
\begin{equation}
  S = \C[p,x_0,x_1,x_2,x_3,x_4],
\end{equation}
with respective grading $(-5,1,1,1,1,1)$. We have a superpotential
\begin{equation}
  \mathcal{W} = p(x_0^5+x_1^5+x_2^5+x_3^5+x_4^5).
\end{equation}
The critical point set of $\mathcal{W}$ is the quintic hypersurface in
$\P^4$ which we denote $X$.  

In the \CY\ phase the Cox ideal is $B_\Sigma=(x_0,\ldots,x_4)$. Let
$\mathsf{w}$ be the $S$-module
\begin{equation}
  \mathsf{w} = \frac{S}{(x_0,\ldots,x_4)}.
\end{equation}
Since $\mathsf{w}$ is annihilated by $\mathcal{W}$ it may also be
viewed as an $S'$-module. We now compute a minimal free $S'$-module
resolution of $\mathsf{w}$:
\begin{equation}
\begin{split}
\xymatrix{\ar[r]&
 {\begin{matrix}S'(-5)\\\oplus\\S'(-3)\{-2\}^{\oplus10}\\\oplus\\S'(-1)\{-4\}^{\oplus5}
\end{matrix}}
 \ar[r]&
 {\begin{matrix}S'(-4)^{\oplus5}\\\oplus\\S'(-2)\{-2\}^{\oplus10}\\\oplus\\S'\{-4\}
 \end{matrix}}
 \ar[r]&
 {\begin{matrix}S'(-3)^{\oplus10}\\\oplus\\S'(-1)\{-2\}^{\oplus5}\end{matrix}}
 \ar[r]&
}\\
\xymatrix@C=25mm{
 {\begin{matrix}S'(-2)^{\oplus10}\\\oplus\\S'\{-2\}\end{matrix}}
 \ar[r]^-{\left(\begin{smallmatrix}-x_1&0&&px_0^4\\x_0&-x_2&&px_1^4\\
   0&x_1&\cdots&px_2^4\\0&0&&px_3^4\\0&0&&px_4^4\end{smallmatrix}\right)}&
 S'(-1)^{\oplus5}\ar[r]^-{\left(\begin{smallmatrix}x_0&x_1&\ldots&x_4\end{smallmatrix}
 \right)}&S'\ar[r]&\mathsf{w}.}
\end{split}
\end{equation}
To write this infinite resolution as a matrix factorization we replace
$S'$ with $S$ and ``roll it up'' following \cite{me:csalg,AG:McExt}.
$\mathsf{w}$ then corresponds to a $16\times16$ matrix factorization:
\begin{equation}
  \xymatrix@R=0mm@C=20mm{
    S(-1)^{\oplus5} &S \\
    \oplus & \oplus\\
    S(-3)[2]^{\oplus10} \ar@<2mm>[r] & S(-2)[2]^{\oplus10}\ar@<2mm>[l]\\
    \oplus & \oplus\\
    S(-5)[4] & S(-4)[4]^{\oplus5}}  \label{eq:16m}
\end{equation}

$T_\Sigma$ in (\ref{eq:comD}) is generated by this matrix
factorizations and all its shifts in both toric and homological
grading.  The category of D-branes on $X$ is given by the category of
matrix factorizations $\mathrm{DGr}S(\mathcal{W})$ quotiented out by
$T_\Sigma$.

From \cite{Gull:Ext,AG:McExt,me:csalg} the category
$\mathrm{DGr}S(\mathcal{W})$ is equivalent to the category of graded
$A$-modules, where
\begin{equation}
  A = \frac{\C[x_0,x_1,\ldots,x_4]}{(x_0^5+x_1^5+\ldots+x_4^5)}.
\end{equation}
Serre's construction of sheaves on a projective variety then shows that
the quotient (\ref{eq:comD}) corresponds to the derived category of
sheaves on the quintic threefold $X$ \cite{Orlov:mfc}.

As an example of this equivalence let us consider the structure sheaf
$\O_X$. This corresponds to the $A$-module $A$ itself. Let us denote
$f=x_0^5+x_1^5+\ldots+x_4^5$. Following the algorithm in
\cite{AG:McExt} we obtain the cyclic free $S'$-resolution
\begin{equation}
\xymatrix{
\ar[r]&
S'\{-4\}\ar[r]^-p&
S'(-5)\{-2\}\ar[r]^-f&
S'\{-2\}\ar[r]^-p&
S'(-5)\ar[r]^-f&
S'\ar[r]&
A} \label{eq:11c}
\end{equation}
and so the structure sheaf $\O_X$ corresponds to the simplest matrix
factorization
\begin{equation}
  \xymatrix@R=0mm@C=20mm{
     S(-5) \ar@<1mm>[r]^-f & S\ar@<1mm>[l]^-p}
      \label{eq:11m}
\end{equation}
Let us use the notation $\mathsf{O}_X$ to refer to this particular
matrix factorization form of $\O_X$. We may, of course, add (using
mapping cones) anything in $T_\Sigma$ to this matrix factorization and
still have a valid representative for $\O_X$. This fact will be used
in section \ref{sss:q2}.

\subsubsection{The monodromy ring}

Let us consider the K-theory and associated monodromy ring in the
compact case. An object in $\DC(\textrm{gr-}S)'$ typically has an
infinite resolution in terms of free $S'$-modules. Following section
\ref{sss:monring} we can associate to any object a power series which 
expresses the associated element of K-theory. As we shall see, it is
very  useful to include an extra variable to express the $R$-grading.

Let $P$ denote the ring of formal power series
\begin{equation}
  \Z[[s_1,s_1^{-1},s_2,s_2^{-1},\ldots,s_r,s_r^{-1},\sigma,\sigma^{-1}]],
\end{equation}
and define a map on free $S'$-modules
\begin{equation}
  k(S'(\mathbf{v})\{m\}) = s_1^{v_1}s_2^{v_3}\ldots s_r^{v_r}\sigma^{-m}.
\end{equation}
By writing free $S'$-module resolutions this extends to a map
\begin{equation}
  k:\DC(\textrm{gr-}S') \to P.
\end{equation}
Ultimately the resolution of any object in $\DC(\textrm{gr-}S')$ is
periodic with period 2 and the product of two consecutive maps, lifted
to an $S$-module map, is of homological degree (and hence $R$-degree since
the two are identified) 2. It
follows, that for any object $\mathsf{a}$ in $\DC(\textrm{gr-}S')$ we
have
\begin{equation}
\begin{split}
  k(\mathsf{a}) &=
  f(s_1,s_2,\ldots,s_r,\sigma)(1+\sigma^2+\sigma^4+\ldots)
    + g(s_1,s_2,\ldots,s_r,\sigma)\\
  &=\frac{f(s_1,s_2,\ldots,s_r,\sigma)}{1-\sigma^2}
    + g(s_1,s_2,\ldots,s_r,\sigma),
\end{split}
\end{equation}
where $f$ and $g$ are finite polynomials. From (\ref{eq:q8}) we see
that $f$ expresses the K-theory class of an object in
$\mathrm{DGr}S(\mathcal{W})$ (with $\sigma$ set equal to one since the
R-grading is equated to the homological grading).

In other words we have
\begin{theorem}
  The map
\begin{equation}
  \kappa:\Obj(\mathrm{DGr}S(\mathcal{W})) \to 
     \Z[s_1,s_1^{-1},s_2,s_2^{-1},\ldots,s_r,s_r^{-1}],
\end{equation}
given by
\begin{equation}
  \kappa(\mathsf{a}) = \frac1{\pi i}\oint_\gamma k(\mathsf{a})\,d\sigma,
\end{equation}
where $\gamma$ is a small loop in the complex plane around $\sigma=1$,
is a group homomorphism 
\begin{equation}
  \kappa:K(\mathrm{DGr}S(\mathcal{W})) \to 
     \Z[s_1,s_1^{-1},s_2,s_2^{-1},\ldots,s_r,s_r^{-1}].
\end{equation}
\end{theorem}
This latter map will be injective but not surjective for all cases
considered in this paper.

From (\ref{eq:comD}) we see that there is a map from the K-theory
associated to D-branes and a quotient of the polynomial
ring $\Z[s_1,s_1^{-1},s_2,s_2^{-1},\ldots,s_r,s_r^{-1}]$.

One may also view K-theory directly in terms of the matrix
factorization (\ref{eq:Pdef}) as the class of $\bar P$ being given as
the class of $P_0$ minus the class of $P_1$. However, since it is
generally easier to compute this in terms of $S'$-modules than matrix
factorizations, the above theorem is more practical.

\subsubsection{The quintic again}  \label{sss:q2}

Referring back to section \ref{sss:quin1} we obtain
\begin{equation}
  \kappa(\mathsf{w}) = (1-s^{-1})^5.
\end{equation}
Since $\mathsf{w}$ generates $T_\Sigma$ in the large radius \CY\
phase, we define our monodromy ring for the quintic \CY\ phase to be
\begin{equation}
  R_0(X) = \frac{\Z[s,s^{-1}]}{\left((1-s^{-1})^5\right)}.
\end{equation}
The subscript zero denotes the large radius phase.
This is not, as a group, isomorphic to the K-theory of $X$. The rank
of $R_0$ is 5 whereas the rank of the K-theory of $X$ is 4.

For most purposes in this paper it will suffice to consider D-brane
charge defined over the {\em rationals}. Algebraic K-theory coincides
with topological K-theory for the quintic (since $h^{1,0}=h^{2,0}=0$)
and topological K-theory over $\Q$ coincides with
$H^{\textrm{even}}(X,\Q)$ via the Chern character isomorphism. So,
given the embedding
\begin{equation}
  i:X\to\P^4,
\end{equation}
we have an isomorphism between the rational K-theory of $X$ and the
pullback, via $i^*$ of the rational K-theory of $\P^4$. The latter is
generated (redundantly) by $i^*\O_{\P^4}(m)=\O_X(m)$ for all $m\in\Z$.

Since $\kappa(\O_X)=1-s^{-5}$, the K-theory class of the D-brane $\O_X$
is represented in $R_0(X)$ by the equivalence class of
$1-s^5$. Similarly $\O_X(m)$ corresponds to $(1-s^{-5})s^m$.
  
Let $M_0^\Q(X)$ denote the rational K-theory of the quintic. Since
monodromy acts on this we may regard $M_0^\Q(X)$ as a $R_0^\Q(X)$-module,
where
\begin{equation}
\begin{split}
  R_0^\Q(X) &= R_0(X)\otimes_\Z\Q\\
    &= \frac{\Q[s,s^{-1}]}{\left((1-s^{-1})^5\right)}.
\end{split} \label{eq:RQq0}
\end{equation}

We therefore have the following
\begin{theorem}
  The rational K-theory of the quintic, $M_0^\Q(X)$, as an
  $\R_0^\Q(X)$-module is given by the image of the map
\begin{equation}
\xymatrix@1{  R_0^\Q(X)\ar[r]^{1-s^{-5}} & R_0^\Q(X).} \label{eq:im1}
\end{equation} 
\end{theorem}
Let $f(s)$ be in the kernel of the map in (\ref{eq:im1}). Thus
\begin{equation}
\begin{split}
  f(s)(s^5-1) &= (s-1)^5g(s)\\
  \hbox{i.e.,}\quad f(s)(s^4+s^3+s^2+s+1) &= (s-1)^4g(s),
\end{split}
\end{equation}
for some Laurent polynomial $g(s)$. Multiplying out any negative
powers of $s$ and using the fact that $\Q[s]$ is a unique
factorization domain, $f(s)$ must have a factor of $(s-1)^4$.
It follows
that $M_0^\Q(X)$ is isomorphic to the cokernel of the 
map\footnote{We
  choose negative powers of $s$ to be consistent with the choice of tilting
  collections such as (\ref{eq:Tq}).}
\begin{equation}
\xymatrix@1@C=18mm{R_0^\Q(X)\ar[r]^{(1-s^{-1})^4} & R_0^\Q(X).}
\end{equation}

Note that as a vector space over $\Q$, $M_0^\Q(X)$ has dimension 4 as
expected. The fact that $M_0^\Q(X)$ is annihilated by $(s-1)^4$ shows that
monodromy around the large radius limit is unipotent of degree 4 as
expected from \cite{Mor:gid}.

The K-theory of $X$ is not given by the cokernel of $(1-s^{-1})^4$
over the integers. In particular, the class of $\O_X$ is 5-divisible
(see (\ref{eq:KO})) in the cokernel of $(1-s^{-1})^4$ but not in $K(X)$.

Now consider the Landau--Ginzburg phase. Here the Cox ideal is
$B_\Sigma=(p)$. The cokernel of the map
\begin{equation}
  p:S(5)\{-2\}\to S,
\end{equation}
is quasi-isomorphic to the complex
\begin{equation}
\xymatrix{
\ar[r]&
S\{-4\}\ar[r]^-p&
S\{-2\}\ar[r]^-f&
S(5)\{-2\}\ar[r]^-p&
S} \label{eq:11d}
\end{equation}
and so is associated
to the matrix factorization (\ref{eq:11m}) (shifted by $[1]$). This
maps under $\kappa$ to $1-s^5$. Our monodromy ring is
then
\begin{equation}
  R^\Q_1(X) = \frac{\Q[s,s^{-1}]}{(1-s^{-5})}.
\end{equation}

We can ``transport'' the D-brane $\O_X$ from the large radius \CY\
phase to the Landau--Ginzburg phase using the trick from
\cite{HHP:linphase,me:toricD}. We use a tilting collection for the
ambient toric variety $V$ of
\begin{equation}
   T = S\oplus S(-1)\oplus\ldots\oplus S(-4). \label{eq:Tq}
\end{equation}
If we write a matrix factorization for a D-brane in the \CY\ phase
using only these summands (which may always be done) then exactly the
same matrix factorization gives the same object in the \LG\ phase.

As written, the matrix factorization $\mathsf{O}_X$ from
(\ref{eq:11m}) representing the 6-brane $\O_X$, is not
in canonical form for transportation as it contains an $S(-5)$. To
remedy this we use a mapping cone from the trivial object
$\mathsf{w}$. To be precise, define $\cX$ from
\begin{equation}
\xymatrix{\mathsf{w}[-4]\ar[rr]^f&&\mathsf{O}_X\ar[dl]\\
&\cX\ar[ul]|{[1]}&}  \label{eq:OO1}
\end{equation}
where $f$ contains the identity map $S(-5)\to S(-5)$ to cancel these
terms away. Thus $\cX$ is the ``correct'' representative for $\O_X$.

Now, in the \LG\ phase, it is $\mathsf{O}_X$ which is trivial while
$\mathsf{w}$ is not. Writing (\ref{eq:OO1}) as
\begin{equation}
\xymatrix{\mathsf{w}[-3]\ar[rr]|{[1]}^(0.6)f&&\mathsf{O}_X\ar[dl]\\
&\cX\ar[ul]&}  \label{eq:OO2}
\end{equation}
we obtain an equivalence between $\cX$ and $\mathsf{w}[-3]$ in the
\LG\ phase. In other words, $\O_X$ is represented by $\mathsf{w}[-3]$
in the \LG\ phase as also argued in \cite{me:lgdict}.

In the monodromy ring $R^\Q_0(X)$, this tilting collection can
be expressed in terms of a Gr\" obner basis. To do this we introduce a
new variable $t=s^{-1}$. In particular if we use {\em lexicographical
  ordering\/} on the ring $\Q[s,t]$ then a Gr\" obner basis of the
ideal $(st-1,(1-t)^5)$ is has leading monomials $t^5$ and $s$. Thus, any
polynomial in the quotient ring (\ref{eq:RQq0}) put in normal form
using this basis will have monomials only in the set
$\{1,t,\ldots,t^4\}$ exactly corresponding to summands of
(\ref{eq:Tq}). The same is true for the \LG\ phase and the
corresponding ideal $(st-1,1-t^5)$.

In particular, the normal form of $1-s^{-5}$, the charge of the
6-brane $\O_X$, is
\begin{equation}
  5t - 10t^2 + 10t^3 - 5t^4.  \label{eq:KO}
\end{equation}

Correspondingly, in the Landau--Ginzburg phase, the charge of
$\mathsf{w}[-3]$ is given by $-(1-t)^5$. This has normal form
with respect to the ideal $(st-1,1-s^5)$ also given by
(\ref{eq:KO}) as we would expect.

In the \LG\ phase the K-theory spanned by the D-brane $\O_X$ and its
images under monodromy around the Gepner point are therefore given by
the image of $(1-s^{-1})^5$. Let $M_1^\Q(X)$ denote this
$R_1^\Q(X)$-module. It follows that $M_1^\Q(X)$ is the cokernel of
\begin{equation}
\xymatrix@1@C=30mm{R_1^\Q(X)\ar[r]^{1+s^{-1}+s^{-2}+s^{-3}+s^{-4}} & R_1^\Q(X).}
\end{equation}
Since this is rank 4 it spans the full K-theory of D-branes in the
\LG\ phase. It also shows that the monodromy about the Gepner 
point is $\Z_5$ and that no periods (linear maps 
$K(X)\to\C$) are invariant under this monodromy,
as can be shown by direct computation of the periods
\cite{CDGP:}.


\section{True Hybrid Model Phases}  \label{s:decomp}

\subsection{The hybrid phase} \label{ss:hybphas}

A simple model with a hybrid phase is a \GLSM\ with one gauge symmetry,
under which the fields $p_0,p_1,x_0,x_1,\ldots,x_5$ have charges
$(-3,-3,1,1,\ldots,1)$ respectively.
The worldsheet superpotential is
\begin{equation}
  \mathcal{W} = p_0f_0(x_i) + p_1f_1(x_i),
\end{equation}
where $f_0(x_i)$ and $f_1(x_i)$ are homogeneous polynomials of degree 3
in $x_0,\ldots,x_5$. We do not need to know any details about
$f_0(x_i)$ and $f_1(x_i)$ except that they are sufficiently generic so
that $X$ is smooth. As an example, one could use \cite{LT:linesCI}
\begin{equation}
\begin{split}
  f_0 &= x_0^3+x_1^3+x_2^3 + x_3x_4x_5\\
  f_1 &= x_3^3+x_4^3+x_5^3 + x_0x_1x_2.
\end{split}  \label{eq:bicub}
\end{equation}

For $r\gg1$
$B_\Sigma = (x_0,x_1,\ldots,x_5)$ and
the low-energy theory is a nonlinear $\sigma$-model on the
intersection $f_0=f_1=0$ in $\P^5$.  This is a typical \CY\ phase. The
non-anomalous $R$-symmetry under which $p_a$ have charge 2
and $x_i$ are invariant is unbroken by the expectation values
of $x_i$ and survives as the superconformal $U(1)_R$ in the low-energy
limit.

For $r\ll-1$ $B_\Sigma = (p_0,p_1)$.  The gauge symmetry is broken to
a $\Z_3$ subgroup by the expectation values of $p_a$.  Classical vacua
are constrained to lie at $x_i=0$ by the $F$-terms, so the space of
classical vacua is $\P^1$ and classically the radius of this is $-r$,
so that na\"\i vely in the limit we have a large-radius, weakly
coupled nonlinear sigma model on this target space.  These vacua are
invariant under the $R$-symmetry to within a gauge transformation.  To
see this explicitly, combine an $R$-rotation with a gauge
transformation such that the $p_a$ are left invariant.  Under this
transformation the $x_i$ all have charge 2/3.  A nonlinear
$\sigma$-model on $\P^1$ cannot be the complete story, of course, and
indeed the fields $x_i$, while obstructed from acquiring nonzero
expectation values, are massless and cannot be integrated away.
Considering fluctuations about a generic classical vacuum, we find
that in addition to the massless fluctuation tangent to $\P^1$ there
are six massless superfields $x_i$ interacting via a cubic
superpotential and transforming under the unbroken discrete gauge
symmetry, describing a Landau-Ginzburg orbifold theory with central
charge $\hat c = 2$.  We can imagine the hybrid theory as a fibration
of a Landau-Ginzburg orbifold over a $\P^1$ base, in analogy with the
construction of \CY\ spaces as fibrations over $\P^1$ with a K3 fibre.

Let us consider the \LG\ theory we find in a particular fibre, i.e.,
fix the value of $p_0$ and $p_1$. We then have some cubic equation in
6 variables. This is a $\hat c=2$ superconformal theory with integral
$U(1)_R$ charges (the $\Z_3$ orbifold projects out all states with
fractional charge) and so must be equivalent \cite{Sei:K3} to a
compactification on either a four-torus or a K3 surface.  In the case
at hand, it is easy to compute that this has 20
deformations of complex structure from deforming the cubic equation
and so must be a rather curious realization of a K3 as noted in
\cite{Vafa:1989china}.

Na\"\i vely one might try to identify this ``K3 surface'' with a cubic
hypersurface in $\P^5$. This correspondence is exactly the
analogue of identifying the mirror of the Z-orbifold as a cubic
equation in $\P^8$ as was done in \cite{Drk:Z,Set:sup}. Probably a
more useful interpretation in the current context is the
following. Since we have defined our conformal field theory in terms
of a superpotential it must be ``algebraic'' in some sense. Since 20
deformations of complex structure are manifest in the deformation of
this defining equation, it must have Picard number $\rho=20-20=0$ (see
\cite{me:lK3}, for example). So the fibre should be thought of as {\em
an algebraic K3 surface with Picard number zero.} 

Even though this description is something of an oxymoron it serves our
model quite well. If one uses such $\rho=0$ K3 surfaces as fibres in a
K3-fibration of a \CY\ threefold then there will be no deformations of
$B+iJ$ associated  with K\"ahler form deformations of the fibre. In
particular there is no deformation that can blow the fibre up into a
geometrical large radius K3 surface. Our model has no true geometric
K3-fibration phase. Instead, the ``K3-fibre'' is stuck at the
$\alpha'$ scale in the exact same way as the mirror of the Z-orbifold
is stuck at some small scale.

This hybrid model was also discussed in
\cite{Caldararu:2007tc}, where the fibres were interpreted as
``noncommutative'' K3s.

It is perhaps worth noting here that the cubic superpotential
degenerates at nine points on $\P^1$.  These do not correspond to
singular conformal theories in the same way that singular fibres
do not imply a singularity in the total space of a fibration.  Here
this is very explicitly clear from the Lagrangian: when the cubic
superpotential is degenerate the pure \LG\ theory acquires a flat
direction.  In the hybrid model, this is lifted by the $F$-terms given
by derivatives of $\mathcal{W}$ with respect to $p_a$ so there is no
decompactification of the space of vacua and no singularity.

\subsection{The monodromy ring} \label{ss:hybmon}

Let us apply the methods of section \ref{s:mon} to this model. The
homogeneous coordinate ring of the ambient toric variety is
\begin{equation}
  S = \C[p_0,p_1,x_0,x_1,x_2,x_3,x_4,x_5],
\end{equation}
with respective grading $(-3,-3,1,1,\ldots,1)$. 
We use the \CY\ $R$-charge assignments, so that both
$p_0$ and $p_1$ have charge 2 while the $x_i$'s all have charge 0.

As described above, this model has two phases and is similar to the
quintic as described in section \ref{s:ccy} in many ways. In the \CY\
phase the Cox ideal is $(x_0,\ldots,x_5)$. Again we define the
$S$-module
\begin{equation}
  \mathsf{w} = \frac{S}{(x_0,\ldots,x_5)},
\end{equation}
which is a no-brane in this phase. In analogy with section
\ref{sss:quin1} (and as computed in \cite{me:modD}) we associate the
following matrix factorization to $\mathsf{w}$:
\begin{equation}
  \xymatrix@R=0mm@C=20mm{
    &S\\
    S(-1)^{\oplus6} &\oplus \\
    \oplus & S(-2)[2]^{\oplus15}\\
    S(-3)[2]^{\oplus20} \ar@<2mm>[r] & *+(20,0){\oplus}\ar@<2mm>[l]\\
    \oplus & S(-4)[4]^{\oplus15}\\
    S(-5)[4]^{\oplus6} & \oplus\\
           & S(-6)[6]}  \label{eq:32m}
\end{equation}

We now have
\begin{equation}
  \kappa(\mathsf{w}) = (1-s^{-1})^6,
\end{equation}
and so the monodromy ring in the \CY\ phase is
\begin{equation}
  R_0^\Q = \frac{\Q[s,s^{-1}]}{\left((1-s^{-1})^6\right)}.
\end{equation}
Thus the sheaf $\O_X$ is associated with the $S$-module
$\sOX$ defined as the cokernel of the map.
\begin{equation}
\xymatrix@1@C=15mm{
  S(-3)^{\oplus 2}\ar[r]^-{f_0,f_1}& S}
\end{equation}
$\sOX$ is an $S'$-module since it is annihilated by $W$.  This can be
repackaged as the matrix factorization \cite{me:modD}:
\begin{equation}
 \xymatrix@R=0mm@C=20mm{
   S(-3)&S(0)\\
   \oplus\ar@<1mm>[r]^{\left(\begin{smallmatrix}f_0&f_1
             \\-p_1&p_0\end{smallmatrix}\right)}
&\oplus\ar@<1mm>[l]^{\left(\begin{smallmatrix}p_0&-f_1
             \\p_1&f_0\end{smallmatrix}\right)}\\
   S(-3)&S(-6)[2]}  \label{eq:Oxmf}
\end{equation}
and
\begin{equation}
  \kappa(\sOX) = (1-s^{-3})^2.
\end{equation}
As for the quintic we define the module $M_0^\Q(X)$ as the image of
$(1-s^{-3})^2$. This is isomorphic to the cokernel of the map
$(1-s^{-1})^4$. Thus we recover maximal unipotency at the large radius
limit as expected.


For the hybrid phase $B_\Sigma=(p_0,p_1)$ and 
it is $\sOX$ that becomes the generating no-brane. 
We then have a monodromy ring
\begin{equation}
  R_1^\Q = \frac{\Q[s,s^{-1}]}{\left((1-s^{-3})^2\right)}.
\end{equation}
Using a tilting object
\begin{equation}
  T = S\oplus S(-1)\oplus\ldots\oplus S(-5),  \label{eq:tiltP33}
\end{equation}
we may transport the D-brane $\O_X$ from the \CY\ phase to the hybrid
phase in a manner identical to the construction for the quintic in
section \ref{sss:q2}. This gives an equivalence
\begin{equation}
  \O_X \sim\mathsf{w}[-3].
\end{equation}
Thus we know that the K-theory of the hybrid phase contains an $R_1^\Q$
given by the image of $\kappa(\mathsf{w}) = (1-s^{-1})^6$. The image
of $(1-s^{-1})^6$ is isomorphic to the cokernel of 
\begin{equation}
(1+s+s^2)^2.  \label{eq:hyb1}
\end{equation}
This is rank 4 and so is the full K-theory for the hybrid phase and
we denote it $M_1^\Q(X)$.

The form of (\ref{eq:hyb1}) immediately yields the monodromy of
periods around
the hybrid limit point. 

We view $M_1^\Q(X)$ as the quotient of $R_1^\Q(X)$ by $(1+s+s^2)^2$
and define the submodule $N\subset M_1^\Q$ as elements of the form
$(1+s+s^2)f(s)$. This is a rank two submodule. Writing
\begin{equation}
  s^3 = 1 - (1-s)(1+s+s^2),
\end{equation}
we see that monodromy on periods three times around the limit point for an
element in $N$ is the identity while the other periods add on
something in $N$. We discuss these periods more in section \ref{s:per}.


\subsection{The symplectic inner product}  \label{s:symp}

The lattice of D-brane charges has a symplectic inner product. For
A-branes on a \CY\ threefold this is nothing more than the intersection
inner product between 3-cycles. For B-branes corresponding to objects
$\cE^\bullet$ and $\cF^\bullet$ in $\DC(X)$ this corresponds to
\begin{equation}
\begin{split}
  \langle\cE^\bullet,\cF^\bullet\rangle &= \sum_i(-1)^i
          \dim\Ext^i(\cE^\bullet,\cF^\bullet)\\
     &=\int_X\ch(\cE^\bullet)^\vee\wedge\ch(\cF^\bullet)\wedge\td(T_X).
\end{split}
\end{equation}

Consider first the linear map from K-theory to $\Z$ given by
Hirzebruch--Riemann--Roch 
\begin{equation}
\begin{split}
  \langle\cE^\bullet\rangle &= \sum_i(-1)^i
          \dim H^i(\cE^\bullet)\\
     &=\int_X\ch(\cE^\bullet)\wedge\td(T_X).
\end{split}
\end{equation}
We may regard this as a linear map from $R(X)$ to $\Z$. This map can
be explicitly determined in the large radius phase since we know
$\O_X(n)$ corresponds to $s^n(1-s^{-3})^2$ in $R(X)$; and we know the
corresponding value in $\Z$. We need only evaluate at $\rank K(X)=4$
points to determine the map fully.
The result is surprisingly simple:
\begin{equation}
  \langle\cE^\bullet\rangle = \Bigl.\mathrm{NF}(\kappa(\cE^\bullet))
                 \Bigr|_{s^0},
\end{equation}
where NF denotes the normal form of the polynomial with respect to the
Gr\"obner basis associated to the tilting collection
(\ref{eq:tiltP33}), i.e., the monomials are of the form
$s^0,s^{-1},\ldots,s^{-5}$. The subscript $s^0$ means take the
coefficient of $s^0$. Now $\ch$ is a {\em ring\/} isomorphism between
$K(X)$ and cohomology, and $\O_X(n)^\vee=\O_X(-n)$. It follows that
\begin{equation}
  \langle\cE^\bullet,\cF^\bullet\rangle =
       \Bigl.\mathrm{NF}\left(\left(\frac{\kappa(\cE^\bullet)}{(1-s^{-3})^2}
               \right)^\vee
                 \kappa(\cF^\bullet)\right)
                 \Bigr|_{s^0},   \label{eq:symp}
\end{equation}
where $f(s)^\vee = f(s^{-1})$. 

The symplectic form is invariant over the moduli space and so we may
compute it in any phase by transporting D-branes back into the large
radius limit phase and using (\ref{eq:symp}). For example, in the
hybrid phase we may choose a basis $1,s^{-1},s^{-2},s^{-3}$ for the
module $M_1^\Q(X)$. With respect to this basis the symplectic product
is given by
\begin{equation}
\begin{pmatrix} \pz0&-6&\pz15&-18\\\pz6&\pz0&-6&\pz15\\
        -15&\pz6&\pz0&-6\\\pz18&-15&\pz6&\pz0\end{pmatrix}.
\end{equation}

\subsection{Periods and special K\"ahler geometry}  \label{s:per}

Until now we have avoided using the period approach to analyzing
monodromy, etc. Instead we can derive all the information we need to
know about the periods from the algebraic methods above. This allows
us to understand issues about the moduli space metric and asymptotic
forms of the D-brane masses without having to solve any differential
equations. 

$M^\Q_i(X)$ are modules over the ring $R^\Q_i(X)$. In the case that
the moduli space is one-dimensional $R^\Q_i(X)$ is a principal ideal
domain. This allows us to use the well-known classification of modules
over P.I.D.'s and, in particular, the notation of rational canonical
form (see, for example, chapter 12 of \cite{DummitFoote:alg}).

Periods are an element of the dual module
\begin{equation}
   Z \in \Hom_{R^\Q_i}(M^\Q_i(X),\C).
\end{equation}
If $M^\Q_i(X)$ is the cokernel of the polynomial $q\in R^\Q_i(X)$ then
monodromy around the limit point expressed as a matrix is simply the
{\em companion matrix\/} for $q$. The asymptotic form of the periods,
can then by simply deduced from the Jordan canonical form of this
matrix.

For example, consider the hybrid phase of the $X$, the bicubic in
$\P^5$. We may put $q=(1+s+s^2)^2$ and then the Jordan canonical form
of the companion matrix is
\begin{equation}
\begin{pmatrix}\omega&1&&\\0&\omega\\&&\omega^2&1\\&&&\omega^2\end{pmatrix}
\end{equation}
Let $z$ be a coordinate on the moduli space such that the limit point
is $z=0$. The asymptotic form of the periods as $z\to0$ are then
\begin{equation}
\begin{split}
  \Phi_1 &= z^{\frac13} + O(z^{\frac43})\\
  \Phi_2 &= \ff1{2\pi i}z^{\frac13}\log(z) + O(z^{\frac13})\\
  \Phi_3 &= z^{\frac23} + O(z^{\frac53})\\
  \Phi_4 &= \ff1{2\pi i}z^{\frac23}\log(z) + O(z^{\frac23})
\end{split} \label{eq:per33}
\end{equation}
It is important to note that the Jordan canonical form cannot be
acquired from the rational canonical form working over $\Q$ --- one
must use complex numbers numbers. This means that the periods of the
pure form (\ref{eq:per33}) are not realized by any actual
D-brane. Instead we only have certain nontrivial linear
combinations. In particular, looking back at the discussion in section
\ref{ss:hybmon} we have any periods associated to objects in $N\subset
M^\Q_1(X)$ going as $z^{\frac13}+O(z^{\frac23})$ and any other period going as
$\ff1{2\pi i}z^{\frac13}\log(z)+O(z^{\frac23}\log(z))$.

Periods give the central charge of a D-brane up to some
normalization. This normalization comes from special K\"ahler
geometry. Let $\alpha_i$ and $\beta^i$ form a basis for $M^\Q_1(X)$
(as a vector space) such that $i$ runs from 1 to half the dimension
and the symplectic form satisfies
\begin{equation}
  \langle \alpha_i,\beta^j \rangle = \delta_i^j.
\end{equation}
Then define the K\"ahler potential \cite{Cand:mir}
\begin{equation}
  K = -\log\left(2\sum_i\Im \left(Z(\alpha_i)\bar Z(\beta^i)\right)\right).
\end{equation}
In the hybrid limit in our example we have
\begin{equation}
  e^{-K} \sim |z|^{\frac23}\log |z|, \quad\hbox{as $z\to0$.}
\end{equation}

The metric on the moduli space is then given by
\begin{equation}
\begin{split}
  g_{z\bar z} &= \partial\bar\partial K\\
              &\sim \frac1{\left(|z|\log|z|\right)^2}\quad\hbox{as $z\to0$.}
\end{split} \label{eq:metric}
\end{equation}
This implies that {\bf the limit point $z=0$ is an infinite distance
  away in the moduli space.} This in turn implies that we should view
this limit as some kind of decompactification.

The mass of a D-brane is given by
\begin{equation}
  m = \frac{e^{\frac K2}|Z|}{g_{10}},
\end{equation}
where $g_{10}$ is the 10-dimensional string coupling. If we go to a
decompactification limit it is natural to hold the 4-dimensional
string coupling, $g_4$, constant. In this case we get
\begin{equation}
  m = \frac1{g_4}e^{\frac K2}|Z|\sqrt{\Vol(X)}. \label{eq:m}
\end{equation}
In the usual decompactification limit of a large radius \CY\ threefold
one then finds using (\ref{eq:m}) that the 0-brane has a mass that is
constant, while 2-branes have mass scaling with the K\"ahler form,
4-branes have mass scaling with the square of the K\"ahler form, etc.

In our hybrid limit we have
\begin{equation}
  e^{\frac K2}|Z(a)| \sim \frac1{\sqrt{\log|z|}},
\end{equation}
for elements $a\in N$, while
\begin{equation}
  e^{\frac K2}|Z(a)| \sim \sqrt{\log|z|},
\end{equation}
for elements $a\in M^\Q_1(X)$, $a\not\in N$.

There is an obvious interpretation of this behaviour in terms of a
decompactification. We assert that $\Vol(X)$ scales as $\log|z|$. Then
we see that elements of $N$ will have masses that are constant in the
limit $z\to0$ and so look like 0-branes. Other elements of $M^\Q_1(X)$
have masses scaling like $\log|z|$ and are 2-branes. Since the volume
of $X$ itself scales like $\log|z|$, it seems that $X$ itself is
2-dimensional. The K\"ahler form on $X$ looks like a ratio of periods:
\begin{equation}
  B+iJ = \frac1{2 \pi i}\log(z)+\ldots,
\end{equation}
as usual.

This is entirely what would expect from the linear $\sigma$-model
picture of the hybrid limit. The base $\P^1$ becomes infinitely large
in the limit.   The interpretation of the hybrid limit in terms of a
fibration is in agreement with this.   Because
the ``K3 fibres'' lack holomorphic curves, we are naturally led to
interpret the instanton expansion about the hybrid limit as 
an expansion in worldsheet instantons wrapping the
base space.  

\begin{figure}
\begin{center}
\includegraphics[angle=270,width=140mm]{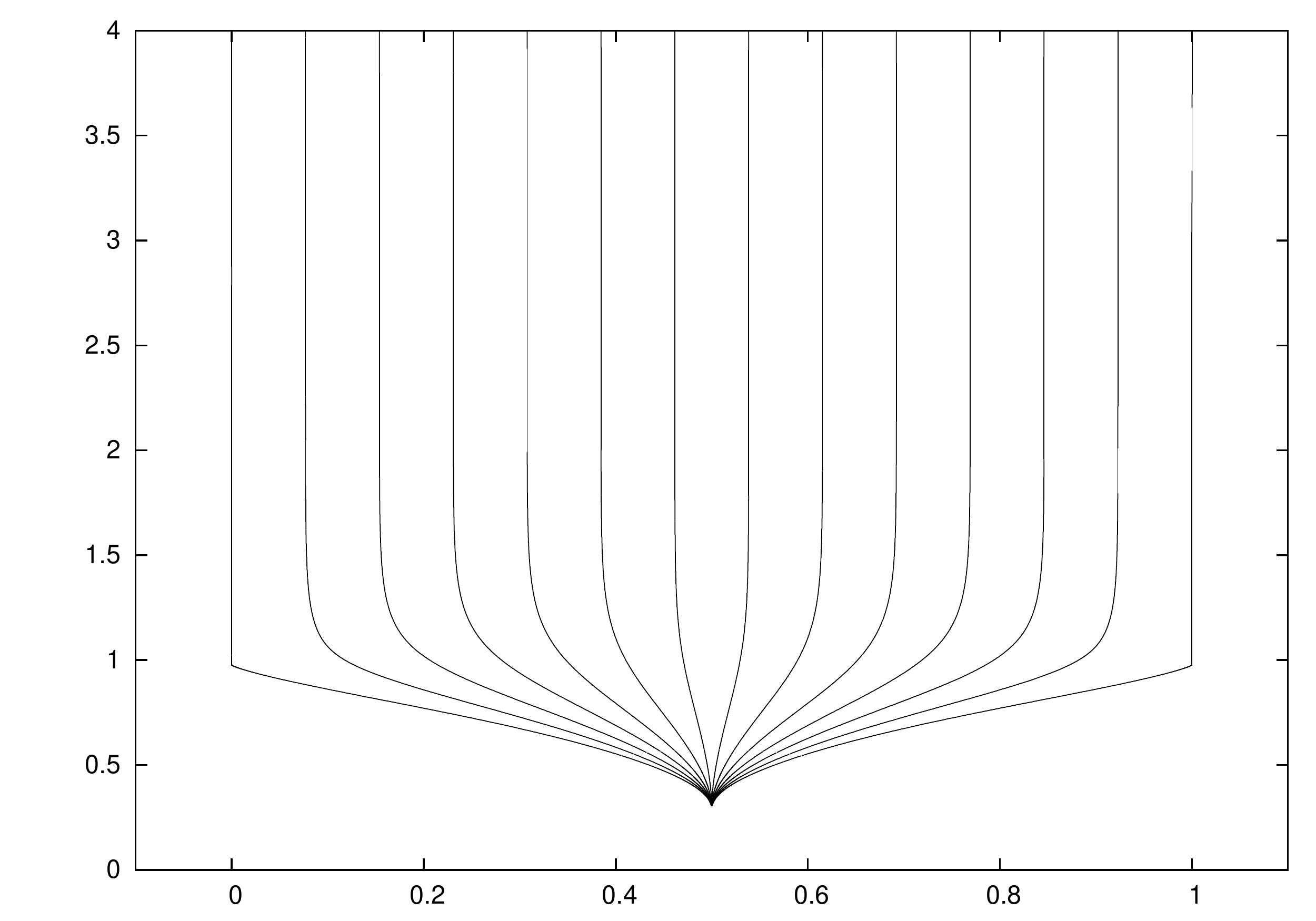}
\end{center}
\caption{Hybrid Moduli Space for $B+iJ$.} \label{fig:hybmod}
\end{figure}

On the other hand, analytic continuation of the worldsheet instanton
series from the \CY\ phase (where they correspond to Gromov-Witten
invariants) as was done in \cite{AGM:sd,me:min-d} might suggest a
different interpretation.  Here one analytically continues the ratio
of periods representing the K\"ahler form in the large radius \CY\
phase. The resulting representation of the moduli space is shown in
figure \ref{fig:hybmod} and is the analogue of the ``scorpion''
diagram of \cite{CDGP:}. The cusp at $B+iJ=\ff12+\ff i{2\sqrt{3}}$
represents the hybrid model limit and it would appear, at first sight,
to represent a \CY\ at {\em finite\/} size. Unfortunately, the true
metric in the moduli space becomes very singular at the cusp and so
figure \ref{fig:hybmod} is misleading.

Let us emphasize that what we are calling ``0-branes'' in the hybrid
limit are not at all the same D-branes that were 0-branes in the large
\CY\ phase limit. The \CY\ 0-branes do not have D-brane charge in $N$
and so become infinitely massive in the hybrid limit. Conversely, the
0-branes of the hybrid model look 6-dimensional in the \CY\ phase
\cite{me:modD} and have divergent masses in the large radius limit.

\subsection{0-Brane probes}  \label{ss:0bp}

Having established the K-theory classes of the candidate 0-branes in
the hybrid limit we would like to explicitly see the corresponding
objects in the category of D-branes. The candidate 0-branes were
proposed in \cite{me:modD} and these turn out to be entirely
consistent with the above.

Define objects
\begin{equation}
  \cP_{[x_0,x_1]} = \Cone\left(\xymatrix@1@C=20mm{
     \mathsf{w}(3)[-2]\ar[r]^-{x_0p_1 - x_1p_0}&\mathsf{w}}\right),
\end{equation}
where $x_0$ and $x_1$ are complex numbers. If $[x_0,x_1]=[0,0]$ then
this object is unstable \cite{me:modD}. It is also easy to prove an
isomorphism
\begin{equation}
  \cP_{[x_0,x_1]} \cong \cP^0_{[\lambda x_0,\lambda x_1]},
\end{equation}
where $\lambda\in\C^*$. Thus the objects $\cP_{[x_0,x_1]}$ are
naturally parametrized by points $[p_0,p_1] = [x_0,x_1]$ on the $\P^1$ of the
hybrid model.

Clearly
\begin{equation}
 \kappa\left(\cP_{[x_0,x_1]}\right) = (1-s^3)\kappa(\mathsf{w}),
\end{equation}
and so $\kappa\left(\cP_{[x_0,x_1]}\right)$ is a multiple of
$(1-s^3)$ in $M_1^\Q(X)$. Thus
$\kappa\left(\cP_{[x_0,x_1]}\right)$ lies in $N$ and has the
right D-brane charge to be a 0-brane from the previous section.

In \cite{me:modD} it was shown that $\cP_{[x_0,x_1]}$ is stable in
the hybrid limit, at least with respect to the obvious decay mode. 

It was also shown in \cite{me:modD} that, in the large radius phase
language, the object $\cP_{[x_0,x_1]}$ corresponds to the complex of
sheaves
\begin{equation}
  \Cone\left(\Omega_{\P^5}^2|_X(3)[1]\to \O_X[3]\right). \label{eq:Plrl}
\end{equation}
Computing the Chern character of $\cP_{[x_0,x_1]}$ in terms of the
cohomology of $X$ yields
\begin{equation}
  \ch(\cP_{[x_0,x_1]}) = 9 + 6x - x^3,
\end{equation}
where $x$ is the generator of $H^2(\P^5;\Z)$ restricted to
$X$. According to the Lefschetz hyperplane theorem, 1 and $x$ generate
$H^0(X;\Z)$ and $H^2(X;\Z)$ respectively. But $x^3$ is Poincar\'e dual
to 9 points on $X$ and thus $x^3/9$ generates $H^6(X;\Z)$. It follows
that $\ch(\cP_{[x_0,x_1]})$ is actually 3-divisible in the lattice of
D-brane charges.

This raises the possibility that $\cP_{[x_0,x_1]}$ is actually a
polystable object which is a direct sum of three objects with the same
D-brane charge. This would imply that $\cP_{[x_0,x_1]}$ is actually
three 0-branes rather than one. We can show this is not the case by
computing $\Hom(\cP_{[x_0,x_1]},\cP_{[x_0,x_1]})$ from
(\ref{eq:Plrl}). Using Macaulay 2 for the form (\ref{eq:bicub}) one
may compute
\begin{equation}
\begin{split}
  \dim\Ext^1(\Omega_{\P^5}^2|_X(3),\O_X) &= 0\\
  \dim\Ext^2(\Omega_{\P^5}^2|_X(3),\O_X) &= 2.
\end{split}
\end{equation}
A standard application of long exact sequences then yields
\begin{equation}
  \dim\Hom(\cP_{[x_0,x_1]},\cP_{[x_0,x_1]}) = 1,
\end{equation}
showing that $\cP_{[x_0,x_1]}$ is {\em not\/} a direct sum of objects.

Since $\cP_{[x_0,x_1]}$ are light in the hybrid limit,
apparently stable and irreducible we therefore assert that they are
indeed the 0-branes that probe the decompactification geometry of the
hybrid limit. From now on we will assume this assertion is true.

Now consider
\begin{equation}
  \cP_{[x_0,x_1]}(n) = \Cone\left(\xymatrix@1@C=20mm{
     \mathsf{w}(n+3)[-2]\ar[r]^-{x_0p_1 - x_1p_0}&\mathsf{w}(n)}\right).
\end{equation}
Monodromy around the hybrid limit point will transform
\begin{equation}
  \cP_{[x_0,x_1]}(n) \to \cP_{[x_0,x_1]}(n+1).
\end{equation}
The D-brane charges of $\cP_{[x_0,x_1]}(n)$ and $\cP_{[x_0,x_1]}(n+1)$ differ.
This means, unlike a conventional large radius limit, the 0-branes
transform into other D-branes under the ``$B$-field'' shift $B\to
B+1$. The objects $\cP_{[x_0,x_1]}(n)$ would appear to have equal right
to be called a 0-brane for any value of $n$.

Since the Cox ideal in the hybrid phase is $(p_0,p_1)$, we have an
exact sequence
\begin{equation}
\xymatrix@1@C=15mm{
  \mathsf{w}(n+6)[-4]\ar[r]^-{\left(
         \begin{smallmatrix}-p_1\\p_0\end{smallmatrix}\right)}&
  \mathsf{w}(n+3)[-2]^{\oplus 2}\ar[r]^-{\left(
         \begin{smallmatrix}p_0&p_1\end{smallmatrix}\right)}&
  \mathsf{w}(n).}
\end{equation}
This implies
\begin{equation}
  \Cone\left(\xymatrix@1@C=15mm{
   \cP_{[x_0,x_1]}(n+3)[-2]\ar[r]^-f&\cP_{[x_0,x_1]}(n)}\right) = 0,
\end{equation}
where $f$ corresponds to any linear combination of $p_0$ and $p_1$
other than $x_0p_1-x_1p_0$. In other words, we have 
\begin{theorem}
There is an isomorphism in the category of D-branes:
\begin{equation}
  \cP_{[x_0,x_1]}(n+3) \cong \cP_{[x_0,x_1]}(n)[2].
\end{equation}  \label{th:mon3}
\end{theorem}

Now $\cP_{[x_0,x_1]}[2]$ and $\cP_{[x_0,x_1]}$ correspond to the
same physical D-brane \cite{Doug:DC,me:TASI-D}. So going three times
around the hybrid limit produces trivial physical monodromy on the
0-branes. Note that we already knew that this was true for D-brane
{\em charge\/} since we observed in section \ref{s:per} that objects
in $N$ were associated to periods of the form $z^{\frac13}$. However, theorem 
\ref{th:mon3} is much stronger since it says that a 0-brane at a given
point on the $\P^1$ comes back to the same 0-brane at the same point.

Theorem \ref{th:mon3} shows that there are three 0-brane-like objects for
each point on $\P^1$. Each of these 0-branes has a different
charge. 


\subsection{Four quadrics on $\P^7$.}   \label{ss:4q}

Consider another model with a one-dimensional moduli space and a
hybrid phase which is very similar to the example above. Let $X$ be
the intersection of four generic quadric polynomials in $\P^7$. This was
studied in \cite{AG:gmi} and more recently in \cite{Caldararu:2007tc}.

The homogeneous coordinate ring is
\begin{equation}
  S = \C[p_0,p_1,p_2,p_3,x_0,\ldots,x_7],
\end{equation}
with respective grading $(-2,-2,-2,-2,1,1,\ldots,1)$. The $R$-charge of
each $p_j$ is set to 2 while the $x_i$'s are all R-charge 0.
The worldsheet superpotential is
\begin{equation}
  \mathcal{W} = p_0f_0(x_i) + \ldots + p_3f_3(x_i),
\end{equation}
where $f_j(x_i)$ are homogeneous polynomials of degree 2
in $x_0,\ldots,x_7$ and are sufficiently generic so
that $X$ is smooth.

The large radius \CY\ phase corresponds to $X$. In the hybrid phase
$B_\Sigma=(p_0,\ldots,p_3)$. The geometrical interpretation of this phase in
terms of the gauged linear $\sigma$-model is that of a $\P^3$, with
homogeneous coordinates $[p_0,\ldots,p_3]$. Over each point in this $\P^3$ we
have a \LG\ theory with fields $x_0,\ldots,x_7$ and a quadric
superpotential. At first sight, a \LG\ theory with quadric
superpotential is completely trivial and so the target space should be
simply $\P^3$. This cannot be, however, since $\P^3$ is not a \CY\
manifold. There are two issues that complicate the statement that the
\LG\ is trivial:
\begin{enumerate}
\item The \LG\ theory is actually a $\Z_2$-orbifolded \LG\ theory.
\item The \LG\ superpotential is not strictly quadric for certain
  points on a surface $L\subset \P^3$.
\end{enumerate}

In \cite{Caldararu:2007tc} it was argued that the proper geometric
interpretation for the hybrid phase, at least in the language of the
gauged linear $\sigma$-model, is that of a double cover of
$\P^3$ branched over $L$. Such a space is a (singular) \CY\
threefold. Will our D-brane probes give the same answer?

The analysis of this case in very similar to the bicubic in $\P^5$:
\begin{itemize}
\item Define
\begin{equation}
   \mathsf{w} = \frac S{(x_0,\ldots,x_7)}.
\end{equation}
The Koszul resolution of this gives a $128\times128$ matrix
factorization in analogy with (\ref{eq:32m}).
\item Define $\sOX$ as the cokernel of the map 
\begin{equation}
\xymatrix@1@C=15mm{
  S(-2)^{\oplus 4}\ar[r]^-{f_0,\ldots,f_3}& S}
\end{equation}
The resulting matrix factorization is an $8\times 8$ factorization
which is an analog of (\ref{eq:Oxmf}).
\item We have monodromy rings:
\begin{equation}
\begin{split}
  R_0^\Q &= \frac{\Q[s,s^{-1}]}{\kappa(\mathsf{w})}\\
   &= \frac{\Q[s,s^{-1}]}{(1-s^{-1})^8},
\end{split}
\end{equation}
and
\begin{equation}
\begin{split}
  R_1^\Q &= \frac{\Q[s,s^{-1}]}{\kappa(\mathsf{\sOX})}\\
   &= \frac{\Q[s,s^{-1}]}{(1-s^{-2})^4}.
\end{split}
\end{equation}
\item We have K-theory modules over the monodromy ring:
\begin{equation}
\begin{split}
  M_0^\Q &= \Image \left(\kappa(\mathsf{\sOX})\right)\\
        &= \coker (1-s^{-1})^4,
\end{split}
\end{equation}
and
\begin{equation}
\begin{split}
  M_1^\Q &= \Image \left(\kappa(\mathsf{\mathsf{w}})\right)\\
        &= \coker (1+s^{-1})^4.
\end{split}
\end{equation}
\item We have a sequence of inclusions of $R_1^\Q$-modules
\begin{equation}
  N_0\subset N_1\subset N_2 \subset N_3\cong M_1^\Q,
\end{equation}
where $N_i$ is the image of $(1+s^{-1})^{3-i}$ in $\coker
(1+s^{-1})^4$. Let $L_i$ be a one-dimensional vector subspace of $N_i$ such
that $N_i=N_{i+1}\oplus L_i$ (and set $L_0=N_0$). An object in $L_i$
has a period that goes as
\begin{equation}
  Z \sim z^{\frac12}\log(z)^i,
\end{equation}
as $z\to0$ in the hybrid limit.
\item The K\"ahler prepotential is given in the hybrid limit by
\begin{equation}
  e^{-K} \sim |z|\left(\log|z|\right)^3,
\end{equation}
yielding the same metric (up to a factor of 3) as (\ref{eq:metric})
and so the hybrid limit is again an infinite distance away in the
moduli space. We therefore again interpret it as a decompactification.
\end{itemize}

Set $B+iJ\sim\ff1{2\pi i}\log(z)$ to simulate the K\"ahler form on
this infinite sized space appearing in the hybrid limit and put
$\Vol(X)\sim\left(\log|z|\right)^3$. The masses of objects in $L_i$
then go as $J^i$ as would befit $2i$-branes on our new space.

Following the construction of 0-branes on $\P^1$ in section
\ref{ss:0bp} we can construct analogous $2i$-branes on $\P^3$ for this
hybrid model. Define
\begin{equation}
  \cP^i_{g_1,g_2,\ldots,g_{3-i}} = \coker\left(\xymatrix@1@C=20mm{
     \mathsf{w}(2)[-2]^{\oplus(3-i)}\ar[r]^-{g_1,g_2,\ldots,g_{3-i}}&
      \mathsf{w}}\right),
\end{equation}
where $g_1,g_2,\ldots,g_{3-i}$ are linearly independent linear
functions in $p_0,\ldots,p_3$. It follows that $\cP^i_{g_1,g_2,\ldots,g_{3-i}}$
is naturally identified with a D-brane wrapping the $2i$-dimensional
subspace of $\P^3$ given by $g_1=g_2=\ldots=g_{3-i}=0$. Note also that
the K-theory class of $\cP^i_{g_1,g_2,\ldots,g_{3-i}}$ is
$(1-s^{-2})^i\kappa(\mathsf{w})$ and so lies in $L_i$. Thus the mass
scales as expected for a $2i$-brane as the $\P^3$ becomes large.

For four linearly independent linear functions $g_1,\ldots,g_4$ one
has in the hybrid phase
\begin{equation}
\coker\left(\xymatrix@1@C=20mm{
     \mathsf{w}(2)[-2]^{\oplus4}\ar[r]^-{g_1,\ldots,g_4}&
      \mathsf{w}}\right) = 0.
\end{equation}
From this, one can deduce
\begin{equation}
\Cone\left(\xymatrix@1{
     \cP^0_{g_1,g_2,g_3}(2)[-2]\ar[r]^-{f}&
      \cP^0_{g_1,g_2,g_3}}\right) = 0,
\end{equation}
for any $f$ linearly independent to $g_1,g_2,g_3$. This gives an
equivalence of 0-branes at the same point
\begin{equation}
  \cP^0_{g_1,g_2,g_3}(2)\cong\cP^0_{g_1,g_2,g_3}[2].
\end{equation}
There are therefore two 0-branes, $\cP^0_{g_1,g_2,g_3}$ and
$\cP^0_{g_1,g_2,g_3}(1)$, for each point in $\P^3$.

We further note that these two 0-branes have opposite K-theory classes
and so the D-brane charges of $\cP^0_{g_1,g_2,g_3}$ and
$\cP^0_{g_1,g_2,g_3}(1)[1]$ coincide. Thus, if one were to try to
compute the moduli space of 0-branes in a particular K-theory class it
would appear to look a lot like a {\em double\/} cover of $\P^3$.

This D-brane interpretation of the geometry of the hybrid limit
appears to be getting very close to the proposals of
\cite{Caldararu:2007tc,Add:P3}. There it was proposed that the relevant
geometry is a double cover of $\P^3$ branched over some surface. The
difference we seem to have with this proposal is that we see no
branching in the double cover.

The 0-branes $\cP^0_{g_1,g_2,g_3}$ are $1024\times 1024$ matrix
factorizations of the superpotential $\mathcal{W}$ where the matrix
entries are polynomials in $p_j$'s and $x_i$'s. In particular, the
matrix factorizations are single-valued functions of $p_j$. In going
around a loop in $\P^3$ it is clearly not the case that a matrix
factorization of the form $\cP^0_{g_1,g_2,g_3}$ can turn into
$\cP^0_{g_1,g_2,g_3}(1)[1]$ in any obvious sense.

Having said that, the geometrical interpretation of
\cite{Caldararu:2007tc,Add:P3} is very appealing and it would be nice to
understand exactly how these apparently different double covers of
$\P^3$ can be reconciled. The discrepancy may be associated with the
fact that our 0-branes $\cP^0_{g_1,g_2,g_3}$ have a K-theory class
which is 2-divisible in the same way that the 0-branes in section
\ref{ss:0bp} were 3-divisible. It would also be interesting to better
understand the relationship between the our 0-branes
$\cP^0_{g_1,g_2,g_3}$ and the spinor bundles of \cite{Add:P3}.


\section{Pseudo-Hybrid Limits}   \label{s:compact}

\subsection{A weighted hybrid}   \label{ss:weighted}

Consider a model with fields $p_0,p_1,p_2,x_0,\ldots,x_6$ with
charges  $(-3,-2,-2,1,1,\ldots,1)$ respectively and a superpotential 
\begin{equation}
  \mathcal{W} = p_0f_0(x_i) + p_1f_1(x_i) + p_2f_2(x_i),
\end{equation}
where $f_0(x_i)$ is a homogeneous polynomial of degree 3 in
$x_0,\ldots,x_7$, and $f_1(x_i)$ and $f_2(x_i)$ are homogeneous
polynomials of degree 2.  For $r\gg 1$, $B_\Sigma =
(x_0,x_1,\ldots,x_6)$ and the low-energy theory is a
nonlinear $\sigma$-model on the \CY\ space $X$ given by the
intersection $f_a=0$ in $\P^6$.  For sufficiently generic choice of
the polynomials this is smooth.  The $R$-symmetry assigns charge 2 to
$p_a$ and leaves $x_i$ invariant.

For $r\ll -1$ we find a new kind of hybrid behavior.  Here
$B_\Sigma=(p_0,p_1,p_2)$.  The superpotential terms then constrain
classical vacua to $x_i=0$ and the space of classical vacua is the
{\em weighted} projective space $\P^2_{\{3,2,2\}}$, with homogeneous
coordinates $[p_0,p_1,p_2]$.  Classically this has radius $-r$ and
in the limit we have a large base.  As we will see, our analysis shows
that this classical expectation is not valid.  

In generic vacua the gauge symmetry is broken completely.  There are
two special loci in the vacuum manifold: the point $P = [1,0,0]$ where
there is an unbroken $\Z_3$, and the curve $C = [0,p_1,p_2]$ where
there is an unbroken $\Z_2$.  The $x_i$ do not acquire masses from the
$D$-terms and interact via a \LG\ superpotential with coefficients
linear in the homogeneous coordinates on the base.  In generic vacua
the \LG\ superpotential is quadratic, the $x_i$ are massive, and the
central charge of the \LG\ fibre is zero.  Related to this is the fact
that generic vacua do not preserve our $R$-symmetry.  The $R$-symmetry
is preserved (up to a gauge transformation) precisely in vacua on the
special loci.  At the point $P$ the preserved symmetry leaves $p_0$
invariant while assigning $R$-charge $2/3$ to all the other fields;
along the curve $C$ the preserved symmetry (leaving $p_1$ and $p_2$
invariant) assigns charge 1 to $x_i$ and charge $-1$ to $p_0$.

These facts tell us that if we can think of the hybrid model as a
fibration, this interpretation cannot be as straightforward as was the
case in our previous models.   If a fibre-wise approximation is valid,
we would expect that generic classical vacua do not lead to
superconformal field theories at low energies.  The special vacua that
do seem to produce such theories lead to theories that are not of the
simple type encountered previously.

We now apply the methods we have described to see whether the
D-brane structure near the limit can help resolve these puzzles.
The homogeneous coordinate ring is
\begin{equation}
  S = \C[p_0,p_1,p_2,x_0,\ldots,x_6],
\end{equation}
with respective grading $(-3,-2,-2,1,1,\ldots,1)$. The $R$-charge of
each $p_j$ is set to 2 while the $x_i$'s are all $R$-charge 0.

The analysis of this case is once more very similar to the bicubic in
$\P^5$, at least at first:
\begin{itemize}
\item Define
\begin{equation}
   \mathsf{w} = \frac S{(x_0,\ldots,x_6)}.
\end{equation}
The Koszul resolution of this gives a $64\times64$ matrix
factorization in analogy with (\ref{eq:32m}).
\item Define $\sOX$ as the cokernel of the map 
\begin{equation}
\xymatrix@1@C=15mm{
  S(-3)\oplus S(-2)^{\oplus 2}\ar[r]^-{f_0,f_1,f_2}& S}
\end{equation}
The resulting matrix factorization is an $4\times 4$ matrix factorization
which is an analog of (\ref{eq:Oxmf}).
\item We have monodromy rings:
\begin{equation}
\begin{split}
  R_0^\Q &= \frac{\Q[s,s^{-1}]}{\kappa(\mathsf{w})}\\
   &= \frac{\Q[s,s^{-1}]}{(1-s^{-1})^7},
\end{split}
\end{equation}
and
\begin{equation}
\begin{split}
  R_1^\Q &= \frac{\Q[s,s^{-1}]}{\kappa(\mathsf{\sOX})}\\
   &= \frac{\Q[s,s^{-1}]}{(1-s^{-3})(1-s^{-2})^2}.
\end{split}
\end{equation}
\item We have K-theory modules over the monodromy ring:
\begin{equation}
\begin{split}
  M_0^\Q &= \Image \left(\kappa(\mathsf{\sOX})\right)\\
        &= \coker (1-s^{-1})^4,
\end{split}
\end{equation}
and
\begin{equation}
\begin{split}
  M_1^\Q &= \Image \left(\kappa(\mathsf{\mathsf{w}})\right)\\
        &= \coker (1+s^{-1}+s^{-2})(1+s^{-1})^2.
\end{split} \label{eq:whc}
\end{equation}
\end{itemize}
The form of (\ref{eq:whc}) makes this case physically very different
to the true hybrids considered above. First consider
\begin{equation}
  P = \Image(1+s^{-1})^2 \subset M_1^\Q.
\end{equation}
This is a two dimensional submodule with monodromy of order three
around the limit $z=0$. Thus we have two periods with a limiting
behavior given by 
$z^{\frac13}$. Next we have
\begin{equation}
  N_0 \subset N_1 \subset M_1^\Q,
\end{equation}
where $N_i=\Image(1+s^{-1}+s^{-2})(1+s^{-1})^{1-i}$. 
As above define $L_0=N_0$ and $L_1$ such that $N_1=L_0\oplus L_1$.
Periods associated to
$L_0$ and $L_1$ must go as $z^{\frac12}$ and $z^{\frac12}\log(z)$
respectively. We also have $M_1^\Q=P\oplus L_0\oplus L_1$.

Computing the symplectic form yields a K\"ahler potential of the form
\begin{equation}
  e^{-K} \sim |z|^{\frac23} + |z|\log|z| + \ldots
\end{equation}
This gives a metric on the moduli space
\begin{equation}
  g_{z\bar z} \sim -|z|^{-\frac53}\log|z|+\ldots,
\end{equation}
which puts the origin $z=0$ at a {\em finite distance\/} away from a
generic point in the moduli space. This limit is not associated to a 
decompactification of the target space.

Since we do not have a decompactification, we do not need to rescale
the D-brane masses by any volume factor to understand their behaviour
near $z=0$. Accordingly we compute $e^{K/2}|Z|$ to see the D-brane
masses.

It follows that the D-branes associated to $N$ have masses that are
nonzero and finite as $z\to0$ while those associated to $L_0$ and
$L_1$ have masses that go to zero. Such massless D-branes are
indicative of a singular conformal field theory
\cite{Str:con}. Indeed, one can compute the normalized Yukawa coupling
as in \cite{CDGP:} and it diverges.

\subsection{Explicit D-branes}  \label{ss:exp}

Let us try to explicitly construct D-branes that are associated with
charges in $P$, $L_0$ and $L_1$.

Copying previous constructions we set
\begin{equation}
\begin{split}
  \cP^0_{[0,x_1,x_2]} &= \coker\left(\xymatrix@1@C=24mm{
     \mathsf{w}(3)[-2]\oplus\mathsf{w}(2)[-2]\ar[r]^-{\left(\begin{smallmatrix}
     p_0&x_2p_1-x_1p_2\end{smallmatrix}\right)}&
     \mathsf{w}}\right)\\
  \cP^0_{[1,0,0]} &= \coker\left(\xymatrix@1@C=15mm{
     \mathsf{w}(2)[-2]^{\oplus2}\ar[r]^-{\left(\begin{smallmatrix}
     p_1&p_2\end{smallmatrix}\right)}&
     \mathsf{w}}\right).\\
\end{split}
\end{equation}
The objects $\cP^0_{[0,x_1,x_2]}$ are 0-branes living on the curve $C$
where we define $C\subset\P^2_{\{3,2,2\}}$ by setting $p_0=0$. These
have charge in class $L_0$ and so all become massless as $z\to0$.

The object $\cP^0_{[1,0,0]}$ is a 0-brane stuck at the point
$[1,0,0]$. It is in the class $P$ and so has nonzero mass in the
hybrid limit.

The new feature arising from the fact that the base in the hybrid is a
weighted projective space is that there is no clear construction of a
0-brane that lives at any point other than on $C$ or
$[1,0,0]$. Classically if one were to write a free resolution of
skyscraper sheaf at an arbitrary point on $\P^2_{\{3,2,2\}}$ then one
would try something like
\begin{equation}
\O_x = \coker\left(\xymatrix@1@C=15mm{
     \O(-6)^{\oplus2}\ar[r]^-{\left(\begin{smallmatrix}
     f_0,f_1\end{smallmatrix}\right)}
     &\O}\right),
\end{equation}
where $f_i$ are of the form $a_0x_0^2+a_1x_1^3+a_2x_2^3$ for complex
$a_0,a_1,a_2$. In the case of the hybrid model we cannot form
combinations like $a_0p_0^2+a_1p_1^3+a_2p_2^3$ because they violate
homogeneity of the $R$-charge grading.

We may also construct 
\begin{equation}
  \cP^1_C = \coker\left(\xymatrix@1{
     \mathsf{w}(3)[-2]\ar[r]^-{p_0}&
     \mathsf{w}}\right)
\end{equation}
which is naturally associated to the 2-brane wrapping the curve
$C$. Its class is in $L_1$ and so it also becomes massless in the
hybrid limit.

\begin{figure}
\begin{center}
\includegraphics[width=50mm]{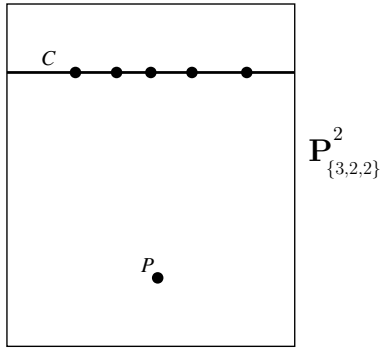}
\end{center}
\caption{Apparent Target Space.} \label{fig:app}
\end{figure}

We have the rather peculiar state of affairs that the 0-brane probe
picture gives something along the lines of figure~\ref{fig:app}. There
is one class of ``points'' that live on the curve $C$ and only one
other isolated point $P$. 

\subsection{Spacetime physics at the limit point}  \label{ss:physw}

What is the physics associated with the hybrid limit for this model?
The set of D-branes that acquires zero mass in this limit corresponds
to a whole $\P^1$'s worth of 0-branes living at point in $C$, the
2-brane wrapping $C$ itself and any stable D-brane whose charge is a linear
combination of these.

To determine the true spectrum of massless particles we need to
quantize these classical BPS objects. In \cite{W:MF} it was argued
that a family of classical D-branes parametrized by $\P^1$ gives rise
to a massless {\em vector\/} in the uncompactified four-dimensional
spacetime. Thus we appear to have an enhanced gauge symmetry. Since
there is only one class of 0-branes going massless, the associated
gauge symmetry should be $\SU(2)$.

It is interesting to compare this situation to the more conventional
picture of nonperturbatively enhanced gauge symmetry appearing in \CY\
compactifications \cite{me:en3g,KMP:enhg}. There one has
\begin{equation}
\xymatrix{E~\ar@{^{(}->}[r]^i\ar@{->>}[d]_q&X\\
Z} \label{eq:EZ}
\end{equation}
where $E$ is a divisor and $Z$ is a curve which we take to be
$\P^1$. The enhanced gauge symmetry occurs when $E$ shrinks down to
the curve $Z$. In this case it is the family of 2-branes wrapping the
fibres of the map $q$ which become massless. Note also that the
4-brane wrapping $E$ also becomes massless and this plays the role of
the magnetic monopole in the $N=2$ $\SU(2)$ theory of \cite{SW:I}.

In \cite{Horj:EZ,AKH:m0} it was argued that the spectrum of massless
D-branes in (\ref{eq:EZ}) is given by (the stable objects in)
$i_*q^*\DC(Z)$. That is, it is essentially the derived category of $Z$
mapped into $\DC(X)$.

Our hybrid model is similar but geometrically quite distinct. We have
points on $C$ and $C$ itself becoming massless. This generates the
whole derived category $\DC(C)$. Thus, again, it is the derived
category of $\P^1$ which is associated with an enhanced gauge
symmetry. But in our case it is the $\P^1$ itself which ``shrinks
down'' and acquires massless wrapped D-branes.

It should also be noted that the theory of enhanced $\SU(2)$ we obtain
is no ordinary $\SU(2)$ theory. The Jordan canonical form of the
monodromy matrix around the hybrid limit coming from (\ref{eq:whc}) is
\begin{equation}
\begin{pmatrix} \omega\\&\omega^2\\
&&-1&\pz1\\&&&-1
\end{pmatrix}.
\end{equation}
The lower block is recognizable as the classical monodromy for
electric and magnetic charges as observed in $\SU(2)$ theories in
\cite{SW:I}. But the monodromy is nontrivial on the other two RR
charges and these are associated to the graviphoton.  Thus the $\SU(2)$
enhanced gauge symmetry is somehow inextricably intertwined with
gravity and is not, therefore, the usual enhanced $\SU(2)$ theory.

Indeed, there is no direction in the moduli space to move where one
might try to ``engineer'' the quantum Seiberg--Witten theory along the
lines of \cite{KV:N=2}. The gauge coupling (or mass scale $\Lambda$)
of the $\SU(2)$ theory is tied to the gravitation coupling and it is
impossible to find any limit where gravity is decoupled from this
$\SU(2)$ theory.


\section{Discussion}   \label{s:discuss}

Gauged linear $\sigma$-models have been a fruitful tool in studying
many aspects of $N$=2 superconformal field theories and the associated
string models.  The gauge theory description is expected to be a
reliable guide to the low-energy physics in the large-$|r|$ limit.  We
expect that in a general model, most of these limits correspond
to some kind of hybrid model.  

In a true hybrid phase we have found that the low-energy dynamics can
usefully be thought of in terms of a fibration.  The fibre is a
conformal field theory of charge $\hat c<3$ and the parameters of this
vary slowly over a large base space of dimension $3-\hat c$.  For the
examples we discuss in the paper the
fibre can be thought of as a \CY\ space of dimension $\hat c$ that is
``stuck'' at small size.  We have given examples of this for $\hat
c=2$ (a K3 with Picard number zero) and $\hat c=0$ (two points).  The
instanton expansion in the gauge theory corresponds to an expansion in
worldsheet instantons wrapping curves in the base (the fibre theory
does not admit field configurations that could contribute to this
expansion) and the limit in which these can be neglected is the limit
in which the base decompactifies.  The analysis of finding light
``0-branes'' reinforces this picture in general.

It is clear, moreover, that true hybrid
limits require a special alignment of the charges in the gauge
theory. In the examples we have considered the key is that the
base $\P^n$ is unweighted in the true hybrid case and weighted in the
pseudo-hybrid case. For example, an intersection of a quadric and a
quartic in $\P^5$ (which was not discussed above) leads to another
pseudo-hybrid phase, since the base is a weighted $\P^1$.
 
We therefore expect that ``most'' limit points will in fact be
pseudo-hybrid phases.  In these phases we have not found a simple
description of the limiting behavior when instanton effects can be
neglected.  The na\"\i ve expectation that the fields describing motion
on the space of classical vacua are weakly interacting is not valid.
Interactions with the other massless fields lead to large corrections
to the kinetic terms and the space of classical vacua does not grow
large in the limit.  Correspondingly, the limit point is at a finite
distance in the moduli space and corresponds to a singular conformal
field theory.

Since the work of Strominger \cite{Str:con} such singularities have
been understood in string theory as associated to the existence of
light D-brane states which become massless in the limit.  If we can
approach the singularity while keeping the volume of the target space
large, we can incorporate these modes into an effective theory while
decoupling gravitational interactions.  This involves taking a limit
in which the string coupling $g\to 0$ and $\alpha'\to 0$ as $z\to 0$
where $z$ is a local coordinate on the moduli space, in such a way
that the effective coupling of the four-dimensional theory remains
finite \cite{KKL:limit}.  In the model of section \ref{s:compact}
there is no way to decouple gravity from the light modes at the
singularity, because in the limit $z\to 0$ one necessarily shrinks
down the whole \CY\ threefold.  In the same way, the singularity
cannot be seen in 5 dimensions from M-theory compactified on the \CY\
threefold.  This is because one must essentially normalize the volume
of the threefold to be one when constructing the M-theory moduli space
of vector multiplet deformations \cite{W:MF}. This normalization
freezes out the deformation we require to reach the singularity.

Obviously further examples will come from considering
cases where the moduli space is dimension greater than one. 
The task of analyzing periods and doing analytic continuation in
multiparameter examples (such as \cite{me:min-d}) can be formidable
but the methods developed in section \ref{s:mon} should simplify this
considerably. 

Finally, there are many interesting computations that can be performed
on our candidate 0-branes which we have not done. First, analysis of
stability was started in \cite{me:modD} but this should be
investigated further. Secondly, there are basic questions about the
moduli space of these 0-branes which remain to be explored. For
example, $\Ext^1(\cP,\cP)$ represents first-order deformations of
D-branes $\cP$ and so gives a na\"\i ve dimension for the moduli
space. One can then test the unobstructedness of such
deformations by using the $A_\infty$-structure. This gives the true
dimension of the moduli space and shows any ``fatness'' of the moduli
space that one might associated with \LG\ theories. Such computations
seem formidable, at first sight, but would be very interesting.

\section*{Acknowledgments}

We wish to thank I.~Melnikov and A.~Roy for useful discussions. The
authors are supported by an NSF grant DMS--0606578.


\begin{thebibliography}{10}

\bibitem{W:phase}
E.~Witten,
\newblock {\em Phases of $N=2$ Theories in Two Dimensions},
\newblock Nucl. Phys. {\bf B403} (1993) 159--222, hep-th/9301042.

\bibitem{AGM:I}
P.~S. Aspinwall, B.~R. Greene, and D.~R. Morrison,
\newblock {\em Multiple Mirror Manifolds and Topology Change in String Theory},
\newblock Phys. Lett. {\bf 303B} (1993) 249--259.

\bibitem{HHP:linphase}
M.~Herbst, K.~Hori, and D.~Page,
\newblock {\em Phases Of $N=2$ Theories In $1+1$ Dimensions With Boundary},
\newblock arXiv:0803.2045.

\bibitem{AG:gmi}
P.~S. Aspinwall and B.~R. Greene,
\newblock {\em On the Geometric Interpretation of $N$ = 2 Superconformal
  Theories},
\newblock Nucl. Phys. {\bf B437} (1995) 205--230, hep-th/9409110.

\bibitem{Caldararu:2007tc}
A.~Caldararu et~al.,
\newblock {\em {Non-Birational Twisted Derived Equivalences in Abelian GLSMs}},
\newblock arXiv:0709.3855.

\bibitem{CDGP:}
P.~Candelas, X.~C. de~la Ossa, P.~S. Green, and L.~Parkes,
\newblock {\em A Pair of Calabi--Yau Manifolds as an Exactly Soluble
  Superconformal Theory},
\newblock Nucl. Phys. {\bf B359} (1991) 21--74.

\bibitem{me:toricD}
P.~S. Aspinwall,
\newblock {\em D-Branes on Toric Calabi--Yau Varieties},
\newblock arXiv:0806.2612.

\bibitem{Cox:}
D.~A. Cox,
\newblock {\em The Homogeneous Coordinate Ring of a Toric Variety},
\newblock J. Algebraic Geom. {\bf 4} (1995) 17--50, alg-geom/9210008.

\bibitem{BO:flop}
A.~Bondal and D.~Orlov,
\newblock {\em Semiorthogonal Decomposition for Algebraic Varieties},
\newblock alg-geom/9506012.

\bibitem{me:TASI-D}
P.~S. Aspinwall,
\newblock {\em D-Branes on Calabi--Yau Manifolds},
\newblock in J.~M. Maldacena, editor, ``Progress in String Theory. TASI 2003
  Lecture Notes'', pages 1--152, World Scientific, 2005,
\newblock hep-th/0403166.

\bibitem{Orlov:mfc}
D.~Orlov,
\newblock {\em Derived Categories of Coherent Sheaves and Triangulated
  Categories of Singularities},
\newblock math.AG/0503632.

\bibitem{me:csalg}
P.~S. Aspinwall,
\newblock {\em Topological D-Branes and Commutative Algebra},
\newblock hep-th/0703279,
\newblock submitted to Communications in Number Theory and Physics.

\bibitem{Eis:mf}
D.~Eisenbud,
\newblock {\em Homological Algebra on a Complete Intersection, with an
  Application to Group Representations},
\newblock Trans. Amer. Math. Soc. {\bf 260} (1980) 35--64.

\bibitem{AG:McExt}
L.~L. Avramov and D.~R. Grayson,
\newblock {\em Resulutions and Cohomology over Complete Intersections},
\newblock in D.~Eisenbud et~al., editors, ``Computations in Algebraic Geometry
  with Macaulay 2'', Algorithms and Computations in Mathematics~{\bf 8}, pages
  131--178, Springer-Verlag, 2001.

\bibitem{Gull:Ext}
T.~H. Gulliksen,
\newblock {\em A Change of Ring Theorem with Applications to Poincare Series
  and and Intersection Multiplicity},
\newblock Math. Scand. {\bf 34} (1974) 167--183.

\bibitem{Mor:gid}
D.~R. Morrison,
\newblock {\em Mirror Symmetry and Rational Curves on Quintic Threefolds: A
  Guide For Mathematicians},
\newblock J. Amer. Math. Soc. {\bf 6} (1993) 223--247, alg-geom/9202004.

\bibitem{me:lgdict}
P.~S. Aspinwall,
\newblock {\em {The Landau--Ginzburg to Calabi--Yau Dictionary for D-Branes}},
\newblock J. Math. Phys. {\bf 48} (2007) 082304, hep-th/0610209.

\bibitem{LT:linesCI}
A.~Libgober and J.~Teitelbaum,
\newblock {\em Lines on Calabi--Yau Complete Intersections, Mirror Symmetry,
  and Picard--Fuchs Equations.},
\newblock Internat. Math. Res. Notices 1993, no. 1, 29--39. (1993) 29--39.

\bibitem{Sei:K3}
N.~Seiberg,
\newblock {\em Observations on the Moduli Space of Superconformal Field
  Theories},
\newblock Nucl. Phys. {\bf B303} (1988) 286--304.

\bibitem{Vafa:1989china}
C.~Vafa,
\newblock {\em {Superstring Vacua}},
\newblock Presented at Symp. on Fields, Strings and Quantum Gravity, Beijing,
  China, May 29 - Jun 10, 1989.

\bibitem{Drk:Z}
P.~Candelas, E.~Derrick, and L.~Parkes,
\newblock {\em Generalized Calabi--Yau Manifolds and the Mirror of a Rigid
  Manifold},
\newblock Nucl. Phys. {\bf B407} (1993) 115--154.

\bibitem{Set:sup}
S.~Sethi,
\newblock {\em Supermanifolds, Rigid Manifolds and Mirror Symmetry},
\newblock Nucl. Phys. {\bf B430} (1994) 31--50, hep-th/9404186.

\bibitem{me:lK3}
P.~S. Aspinwall,
\newblock {\em K3 Surfaces and String Duality},
\newblock in C.~Efthimiou and B.~Greene, editors, ``Fields, Strings and
  Duality, TASI 1996'', pages 421--540, World Scientific, 1997,
\newblock hep-th/9611137.

\bibitem{me:modD}
P.~S. Aspinwall,
\newblock {\em Probing Geometry with Stability Conditions},
\newblock arXiv:0905.3137.

\bibitem{DummitFoote:alg}
D.~S. Dummit and R.~M. Foote,
\newblock {\em Abstract Algebra},
\newblock Prentice Hall, 2003.

\bibitem{Cand:mir}
P.~Candelas and X.~C. de~la Ossa,
\newblock {\em Moduli Space of Calabi--Yau Manifolds},
\newblock Nucl. Phys. {\bf B355} (1991) 455--481.

\bibitem{AGM:sd}
P.~S. Aspinwall, B.~R. Greene, and D.~R. Morrison,
\newblock {\em Measuring Small Distances in $N=2$ Sigma Models},
\newblock Nucl. Phys. {\bf B420} (1994) 184--242, hep-th/9311042.

\bibitem{me:min-d}
P.~S. Aspinwall,
\newblock {\em Minimum Distances in Non-Trivial String Target Spaces},
\newblock Nucl. Phys. {\bf B431} (1994) 78--96, hep-th/9404060.

\bibitem{Doug:DC}
M.~R. Douglas,
\newblock {\em D-Branes, Categories and $N$=1 Supersymmetry},
\newblock J. Math. Phys. {\bf 42} (2001) 2818--2843, hep-th/0011017.

\bibitem{Add:P3}
N.~Addington,
\newblock {\em The Derived Category of the Intersection of Four Quadrics},
\newblock arXiv:\allowbreak0904.1764.

\bibitem{Str:con}
A.~Strominger,
\newblock {\em Massless Black Holes and Conifolds in String Theory},
\newblock Nucl. Phys. {\bf B451} (1995) 96--108, hep-th/9504090.

\bibitem{W:MF}
E.~Witten,
\newblock {\em Phase Transitions in M-Theory and F-Theory},
\newblock Nucl. Phys. {\bf B471} (1996) 195--216, hep-th/9603150.

\bibitem{me:en3g}
P.~S. Aspinwall,
\newblock {\em Enhanced Gauge Symmetries and Calabi--Yau Threefolds},
\newblock Phys. Lett. {\bf B371} (1996) 231--237, hep-th/9511171.

\bibitem{KMP:enhg}
S.~Katz, D.~R. Morrison, and M.~R. Plesser,
\newblock {\em Enhanced Gauge Symmetry in Type II String Theory},
\newblock Nucl. Phys. {\bf B477} (1996) 105--140, hep-th/9601108.

\bibitem{SW:I}
N.~Seiberg and E.~Witten,
\newblock {\em Electric - Magnetic Duality, Monopole Condensation, and
  Confinement in N=2 Supersymmetric Yang-Mills Theory},
\newblock Nucl. Phys. {\bf B426} (1994) 19--52, hep-th/9407087,
\newblock (erratum-ibid. {\bf B430} (1994) 485-486).

\bibitem{Horj:EZ}
R.~P. Horja,
\newblock {\em Derived Category Automorphisms from Mirror Symmetry},
\newblock math.AG/\-0103231.

\bibitem{AKH:m0}
P.~S. Aspinwall, R.~L. Karp, and R.~P. Horja,
\newblock {\em Massless D-branes on Calabi-Yau threefolds and monodromy},
\newblock Commun. Math. Phys. {\bf 259} (2005) 45--69, hep-th/0209161.

\bibitem{KV:N=2}
S.~Kachru and C.~Vafa,
\newblock {\em Exact Results For N=2 Compactifications of Heterotic Strings},
\newblock Nucl. Phys. {\bf B450} (1995) 69--89, hep-th/9505105.

\bibitem{KKL:limit}
S.~Kachru et~al.,
\newblock {\em Nonperturbative Results on the Point Particle Limit of N=2
  Heterotic String Compactifications},
\newblock Nucl. Phys. {\bf B459} (1996) 537--558, hep-th/9508155.

\end{thebibliography}

\end{document}